\documentclass[aps,pre,twocolumn,superscriptaddress]{revtex4-1}

\usepackage{amsmath}
\usepackage{amssymb}
\usepackage{graphicx}
\usepackage{color}
\usepackage{hyperref}

\usepackage[labelfont=bf,labelsep=period,justification=raggedright]{caption}

\newcommand{\avg}[1]{\left < #1 \right >}
\newcommand{\br}[1]{\left ( #1\right )}
\newcommand{\abs}[1]{\left |{#1}\right |}
\newcommand{\sqbr}[1]{\left [ #1\right ]}

\newcommand{\andd}{\text{ and }}
\newcommand{\diff}[2]{\frac{d #1}{d #2}}
\newcommand{\pdiff}[2]{\frac{\pd #1}{\pd #2}}

\newcommand{\avgabssq}[1]{\avg{\abs{#1}^2}}
\newcommand{\vect}[1]{\boldsymbol{#1}}
\newcommand{\mat}[1]{\mathbf{#1}}
\newcommand{\pd}{\partial}
\newcommand{\GNR}{\textit{GNR}}
\newcommand{\ds}{\displaystyle}
\newcommand{\AMOLF}{ FOM Institute for Atomic and Molecular Physics, Science
Park 104, 1098 XG, Amsterdam}
\newlength{\colwidth}
\newcounter{thebox}
\newcommand{\letter}[1]{
\begin{minipage}[t]{0.01\colwidth}
{\fontfamily{phv}\selectfont\huge #1}
\end{minipage}
}
\definecolor{LightBlue}{rgb}{.8,0.8470,.9019}

\begin{document}

\title{The effect of feedback on the fidelity of information transmission of time-varying signals}
\date{\today}
\author{Wiet Hendrik de Ronde}
\email{deronde@amolf.nl}
\author{Filipe Tostevin}
\author{Pieter-Rein ten Wolde}
\affiliation{\AMOLF}

\begin{abstract}
Living cells are continually exposed to environmental signals that vary in time. These signals are
detected and processed by biochemical networks, which are often highly stochastic. To understand how
cells cope with a fluctuating environment, we therefore have to understand how reliably biochemical
networks can transmit time-varying signals. To this end, we must understand both the noise
characteristics and the amplification properties of networks. In this manuscript, we use information
theory to study how reliably signalling cascades employing autoregulation and feedback can transmit
time-varying signals. We calculate the frequency-dependence of the gain-to-noise ratio, which
reflects how reliably a network transmits signals at different frequencies.
We find that the gain-to-noise ratio may differ qualitatively from the power spectrum of the output,
showing that the latter does not directly reflect signaling performance. Moreover, we find that
auto-activation and auto-repression increase and decrease the gain-to-noise ratio for all of
frequencies, respectively. Positive feedback specifically enhances information transmission at low
frequencies, while negative feedback increases signal fidelity at high frequencies. Our analysis not
only elucidates the role of autoregulation and feedback in naturally-occurring biological networks,
but also reveals design principles that can be used for the reliable transmission of time-varying
signals in synthetic gene circuits.
\end{abstract}

\maketitle

\section{Introduction}

Living cells constantly have to respond and adapt to a changing environment. In
some cases, such as in response to a changing sugar concentration
\cite{bennett08}, a cell may wish to integrate out rapid variations and
only respond to slow variations of the environmental signal, while in other
cases, such as osmo adaptation \cite{mettetal08} or bacterial chemotaxis
\cite{tu08}, the cell needs to do the opposite --- respond to rapid but not slow variations
(adaptation). Indeed, to understand how cells cope with a
fluctuating environment, we have to understand how cells transduce time-varying
signals. Cells detect, process, and transduce signals via biochemical networks,
which are the information processing devices of life. However, experiments in
recent years have demonstrated that biochemical networks are often highly
stochastic \cite{elowitz02,swain02}. This raises the question how reliably
biochemical networks can transmit time-varying signals in the presence of noise.

Interestingly, biochemical networks exploit commonly recurring architectures
\cite{ma'ayan05, milo02}, such as autoregulation, cascades, and feedback, to
process signals. These network motifs often implement signal amplification in
order to raise the level of the input signal relative to the noise.
Amplification can be characterised by the {\em gain}, the fold-change in the
signal amplitude. However, it is important to recognise that such amplification
can not only increase the levels of the desired signal, but can also amplify the
noise itself. Therefore, to understand the possibilities and limitations of
different network motifs for enhancing the fidelity of signal transduction, we
need to understand how both the signal and the noise are propagated through
these motifs. Specifically, information theory indicates that the reliability of
signal transmission is determined by the ratio of the gain of the network to the
total noise in the output signal --- the gain-to-noise ratio. Moreover, to
assess how reliably signals of different temporal characteristics are
transduced, we have to understand the frequency dependence of the gain and the
noise. Importantly, we expect that different network architectures will affect
the frequency-dependence of the gain and the noise differently, which means that
we have to study both these quantities. In this manuscript, we study the
frequency-dependence of the gain-to-noise ratio for simple cascades, and for
cascades employing autoregulation and feedback. This allows us to elucidate how
autoregulation and feedback can shape the frequency range over which signals can
be transduced reliably.

Information theory provides a formalism for quantifying the reliability of
information transmission in the presence of noise \cite{shannon48}. A natural
measure for the fidelity of signal transmission from an input signal $S$ to an
output signal $X$ (the network response) is the mutual information between $S$ and $X$, which is
defined as
\begin{multline}
  I(S,X)=H\br{S}-H\br{S\vert X}\\=-\int dS\, p\br{S}\log\sqbr{p\br{S}} -\\
  \br{-\int dX\, p\br{X}\int dS\, p\br{S\vert X}\log\sqbr{p\br{S\vert X}} }.
\end{multline}

Here, $p(S)$ and $p(X)$ are the probability distributions of possible input and
output signals respectively, and $p\br{S\vert X}$ is the conditional probability
of $S$ once $X$ is specified. The mutual information quantifies the reduction in
entropy of (or uncertainty about) the signal after one obtains knowledge of the
network response, averaged over all possible responses. In other words, $I(S,X)$
is how much we learn (on average) about $S$ by measuring $X$. For a
deterministic system, every $S$ leads to a unique $X$ (we assume no degeneracy).
Measuring $X$ thus precisely specifies $S$, such that the uncertainty in $S$
after a measurement of $X$ is $H(S|X)=0$ and $I\br{S,X}=H\br{S}$. However, in the presence of noise
in the network each input $S$ will lead to a distribution of possible
outputs $X$. As a result, an observed $X$ can correspond to multiple $S$ values
and $I\br{S,X}\le H\br{S}$. For completely uncorrelated $S$ and $X$, $I(S,X)=0$.
By construction, the mutual information is symmetric, such that
$I\br{S,X}=I\br{X,S}$.

Recently, the mutual information has been used to study the reliability of
information transmission in biochemical networks \cite{ziv07, tkacik08,
walczak09, mehta09}. However, these studies considered only the steady-state
response of a network to a distribution of \textit{constant} input signals,
which do not change on the timescale of the network response. Yet, in many
biological systems, it cannot be assumed that the input signal is constant on
the timescale of the network response.

Indeed, in many systems the message is encoded in the {\em temporal dynamics} of
the input signal. A well-known example is bacterial chemotaxis, where the
concentration of the intracellular messenger protein depends not on the steady-state ligand
concentration, but rather on the change of this concentration in the recent past
\cite{segall83} --- the response of the network thus
depends on the history of the input signal. Moreover, the extracellular signal
may be encoded in the temporal dynamics of the intracellular signal transduction
pathway. An interesting example is provided by the rat PC-12 system: while
stimulation with a neuronal growth factor gives rise to a sustained response of
the Raf-Mek-Erk pathway, stimulation by an epidermal growth factor gives rise to
a transient response of this pathway \cite{marshall95}. In all these cases, the
message is encoded not in the concentration of some chemical species at a given
moment in time, but rather in its concentration as a function of time. This
means that to understand how reliably the network can transmit information, we
need to know how accurately an input signal as a function of time --- the input
{\em trajectory} $s\br{t}$ --- can be mapped onto an output trajectory $x\br{t}$. We thus need
to understand the mutual information
between the two trajectories, $I\br{s\br{t},x\br{t}}$.

The ability of a biochemical network to transduce a time-varying input signal
depends on the correlation time of the input signal and the architecture and
response dynamics of the network. An instructive example is provided by the
chemotaxis network of the bacterium {\em Escherichia coli}. This network employs
integral negative feedback \cite{yi00}, as a result of which the intracellular
messenger protein can adapt to a constant extracellular ligand concentration. This means that
the signalling network cannot respond to changes in ligand concentration that
occur on time scales longer than the adaptation time. At the other end of the
frequency spectrum, changes in the messenger protein that occur on time scales
shorter than the motor switching time will be integrated out; indeed, the
network cannot respond reliably to rapidly varying input signals
\cite{tostevin09}. The architecture and the response dynamics of the processing
network thus determines the frequency range over which signals can be transduced
reliably.

Recently, we have applied information theory to biochemical networks and studied
the mutual information between in- and output trajectories, $I\br{s\br{t},x\br{t}}$
\cite{tostevin09}.
Here, we apply this framework to study the propagation of time-varying signals
through a number of network motifs---cascades, autoregulation, and feedback. It
is known that for \textit{constant} signals (or, to be more precise, signals
that do not vary on the time scale of the network response time), the mutual
information decreases as a function of cascade length \cite{walczak09}. The same
also holds true for time-varying signals. Indeed, the data-processing inequality
states that in a cascade with $n$ nodes, the information about the input encoded
in the signal at node $i+1$ cannot be greater than the information at node $i$.
Once lost, information about the input cannot be recovered later in the cascade.
Simply increasing the length of a signalling cascade therefore can never
increase the transmitted information. Conversely, maximising the total
transmitted information cannot be the driving force behind the evolution of such
cascades.

Cascades, however, often employ autoregulation and feedback, which can be used
to shape the response of the network to signals of different frequencies.
Importantly, autoregulation and feedback affect not only the
frequency-dependent gain, which describes how strongly an input signal at a
particular frequency is amplified in the absence of any biochemical noise, but
also the frequency-dependence of the noise. While the frequency-dependence of
the gain \cite{samoilov02, ingalls04, locasale08} and the noise \cite{simpson03}
have been studied separately, the frequency-dependence of their ratio, the
gain-to-noise ratio, has not. However, it is the gain-to-noise ratio which
determines how reliably an input signal at a particular frequency can be
transmitted \cite{tostevin09}. In fact, as we will show, autoregulation and
feedback affect the frequency-dependence of the gain and the noise differently,
which means that it is essential to study these quantities together.

In this manuscript, we study the frequency-dependent gain-to-noise ratio using a
Gaussian model. In the next section, we describe this model, and how we can use
it to compute the frequency-dependent gain-to-noise ratio and the information
transmission rate, which is given by the integral of this ratio over all
frequencies \cite{tostevin09}. In section {\em Results} we discuss the
frequency-dependent gain-to-noise ratio of simple cascades, and cascades
employing feedback and autoregulation. Our results highlight the idea
that the output power spectrum is not a direct measure for the information content of the
output signal---the output power spectrum can differ {\em qualitatively} from
the spectrum of the gain-to-noise ratio. We also show (Fig.~\ref{fig:cartoon}) that positive
regulation tends to increase the gain-to-noise ratio, while negative regulation tends to
decrease it. Moreover, we show that the frequency spectra of motifs with negative feedback can
exhibit windows in which the gain-to-noise ratio is increased; these motifs can
thus act as band-pass filters for information transmission. Finally, we discuss
some of the implications of our findings and the limitations of our analysis.

\section{Methods}

We consider information transmission through a biochemical network from an input signal $s\br{t}$ to
an output signal $x\br{t}$. The dynamics of the network can be described mathematically by a set of
coupled Langevin equations \cite{gillespie00} for the signal, response and an arbitrary number of
intermediate components $v_i$, in vector form $\vect{v}$. In using the Langevin representation we
assume that the copy number of each component is large such that the discrete number of molecules
can be approximated by a continuous concentration.
\begin{subequations} \label{eqn:langevin_sys}
  \begin{equation} \label{eqn:langevin_sys_a}
    \frac{ds}{dt}=f^{+}_s(s)+f^{-}_s(s)+\Gamma\br{t}
\end{equation}
\begin{equation}
  \frac{d\vect{v}}{dt}=\vect{f}_v^{+}\br{s,\vect{v},x}\\
  -\vect{f}_v^{-}\br{s,\vect{v},x}+\vect{\eta}_v\br{t}
\end{equation}
\begin{equation}
  \frac{dx}{dt}=f_x^{+}\br{s,\vect{v},x}\\
  -f_x^{-}\br{s,\vect{v},x}+\eta_x\br{t}.
\end{equation} \end{subequations}
Here, $f_i^{+}$ and $f_i^{-}$ contain all the reactions involving the production
and degradation of component $i$, respectively. $f_s^{+}$ and $f_s^{-}$ only depend on $s$, so
that we restrict our analysis to networks that do not feed back onto $s$ itself. In these cases, the
gain-to-noise ratio is independent of the input signal \cite{tanasenicola06}, as discussed in more
detail below Eqn.~\ref{eqn:rate_SN}. $\Gamma\br{t}$ is a stochastic driving process that serves
to define the ensemble of possible input signals. The various noise sources $\eta_i$ are taken to be
independent and Gaussian-distributed \cite{paulsson04, tanasenicola06, scott07},
such that $\avg{\eta_i\br{t}\eta_j\br{t'}}=\avgabssq{\eta_i}\delta_{ij}\delta(t-t')$.
Here, we note that the assumption of independent noise sources is only made to simplify the
analysis. (Anti)-correlations between noise sources can affect noise propagation
\cite{tanasenicola06}, and can be included by a straightforward extension of the present
discussion. Furthermore, we assume that $\avgabssq{\eta_i}=\avg{f_i^+}+\avg{f_i^-} =2\avg{f_i^+}$
and $\avgabssq{\Gamma}=2\avg{f_s^{+}}$ \cite{warren06}, the sum of the production and degradation
terms.

\begin{figure} [!ht]
\includegraphics[]{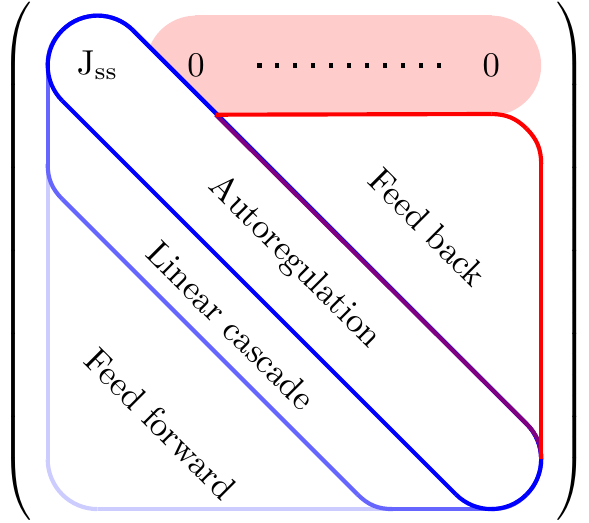}
	\caption{{\bf The Jacobian matrix}. The entries of the Jacobian
		matrix encode the structure of the reaction network.\label{fig:mat_jac} }
\end{figure}

We introduce the vector $\vect{y}=\br{s;\vect{v};x}$ and
$\vect{\eta}=\br{\Gamma,\vect{\eta}_v,\eta_x}$ and assume the
network has a steady state $\avg{\vect{y}}$. Linearizing around this steady
state we obtain
\begin{equation} \label{eqn:lin_sys}
	\frac{d\vect{\tilde{y}}}{dt}=\mat{J}\vert_{\vect{y}=\avg{\vect{y}}}
		\vect{\tilde{y}} +\vect{\eta}.
\end{equation}
Here $\tilde{y_i}=y_i-\avg{y_i}$ is the deviation of the concentration of component $i$
from its steady-state value, $\avg{y_i}$, and $\mat{J}$ is the Jacobian
evaluated at the steady state \footnote{
(We note that with the chosen expression for the size of the random
	events $\avg{\eta_i\eta_j}$, the linearized Langevin equations lead to the
	Fluctuation-Dissipation theorem or Linear Noise Approximation \cite{vankampen,
	paulsson05})}.
$J_{ij}$ describes the response of the component $i$ to small changes in
component $j$, while keeping all other components at their steady-state levels.
The diagonal element $J_{ii}=-\tau_i^{-1}$ is the relaxation time or dissipative
time scale of component $i$; it describes the time scale on which component $i$
relaxes back to its steady-state value after a perturbation. After
linearization, the architecture of the network is encoded in the structure of
the Jacobian matrix (see Fig.~\ref{fig:mat_jac}): the diagonal terms correspond
to autoregulation, the lower triangular part to downstream (feedforward)
regulation and the upper triangular part to upstream (feedback) regulation.
Since we restrict ourselves to systems without feedback from the network to the
signal itself, we require that all elements on the first row of $\mat{J}$ are
zero but for $J_{ss}$.

We take as our input signal the variations $\tilde{s}$. A linear system does not
change the frequency of the transmitted signal, but only the amplitude and the
phase. Since Eqn.~(\ref{eqn:lin_sys}) is linear in $\vect{\tilde{y}}$, we can
calculate exactly the power spectra of the network components \cite{warren06},
\begin{equation}
	\mat{P}=\sqbr{i\omega\mat{I}-\mat{J}}^{-1}\mat{\Xi}
		\sqbr{-i\omega\mat{I}-\mat{J}^T}^{-1},
\end{equation}
where $P_{ij}(\omega)=\avg{\tilde{Y}_i(\omega)\tilde{Y}_j(-\omega)}$ is the
(cross-)power spectrum of $\tilde{y}_i$ and $\tilde{y}_j$, $\tilde{Y}_i\br{\omega}$ is the Fourier
transform of $\tilde{y}_i\br{t}$, $\mat{I}$ is the identity matrix, and
$\mat{\Xi}$ is the noise matrix with entries
$\Xi_{ij}=\avg{\eta_i(\omega)\eta_j(-\omega)}=\avgabssq{\eta_i}\delta_{ij}$. The
power spectrum is a commonly used tool to study time-varying signals, and
describes how the total power of a signal is distributed over different
frequencies. Power at low frequencies is related to slow variations of the
signal, while power at high frequencies corresponds to rapid fluctuations. The
integral of the power spectrum over all frequencies equals the total variance of
the signal.

The information transmission rate for time-varying signals is \cite{borst99,munakata06}
\begin{multline} \label{eqn:rate_transmission}
	\lim_{T\to\infty}\frac{I\br{s\br{t},x\br{t}}}{T}=R\br{s\br{t},x\br{t}}\\=-\frac{1}{2\pi}
 \int_0^\infty d\omega
		\ln\sqbr{1-\Phi_{sx}(\omega)},
\end{multline}
where $T$ is the length of the trajectory and $\Phi_{sx}(\omega)$ is the coherence function,
defined as
\begin{equation} \label{eqn:coh_function}
	\Phi_{sx}(\omega)=\frac{\abs{P_{sx}(\omega)}^2}{P_{ss}(\omega)
		P_{xx}(\omega)}.
\end{equation}
$\Phi_{sx}(\omega)$ is a measure of the average correlation between
the in- and output signals in the frequency domain. For completely independent in- and output
signals, $\Phi_{sx}(\omega)=0$, while for a noiseless system $\Phi_{sx}(\omega)=1$.

The power spectrum of the output signal, $P_{xx}(\omega)$, can be decomposed as
\begin{eqnarray}
P_{xx}(\omega)&\equiv& \Sigma(\omega)+N(\omega),\\
&\equiv&g^2(\omega)P_{ss}(\omega) + N(\omega).\label{eqn:Sadd}
\end{eqnarray}
Here, $\Sigma(\omega)\equiv g^2\br{\omega} P_{ss}(\omega)$ is the transmitted signal,
$g^2(\omega)\equiv \abs{P_{sx}(\omega)}^2/P_{ss}^2(\omega)$ is the frequency-dependent
gain,  $P_{ss}(\omega)$ is the power spectrum of the input signal and $N(\omega)$ is the
frequency-dependent noise. With these definitions,
the coherence function, Eq. \ref{eqn:coh_function}, can be recast as
\begin{equation} \label{eqn:basic_eqn}
	\Phi_{sx}(\omega)=\frac{\Sigma(\omega)}
		{N(\omega)+\Sigma(\omega)},
\end{equation}
and the mutual information rate can be rewritten as \cite{tostevin09}
\begin{equation} \label{eqn:rate_SN}
	R\br{s\br{t},x\br{t}}=\frac{1}{2\pi}\int_0^\infty d\omega
		\ln\sqbr{1+\frac{g^2(\omega)}{N(\omega)}P_{ss}(\omega)}.
\end{equation}
We see that the information transmission rate depends on the power spectrum of
the input signal, $P_{ss}(\omega)$, and on the gain-to-noise ratio
$g^2(\omega)/N(\omega)$.

As discussed in Ref. \cite{tanasenicola06}, in a biological system the reaction
that detects the input signal can, depending on the nature of the detection
reaction, introduce significant correlations between the variations in the input
signal and the intrinsic noise of the reactions that constitute the processing
network. These correlations are a consequence of the molecular character of the
components and are thus unique to biochemical networks. If the detection
reaction does not introduce correlations, then Eqn. \ref{eqn:Sadd} is the
spectral-addition rule \cite{tanasenicola06}. The noise $N(\omega)$ is then the
intrinsic noise of the processing network and also $g^2(\omega)$ only depends on
properties of the processing network. On the other hand, if the detection reaction does introduce
correlations, then the output power spectrum $P_{xx}(\omega)$ can be written in
the form of Eqn. \ref{eqn:Sadd}, but then $N(\omega)$ and $g^2(\omega)$ depend
not only on characteristics of the processing network, but also on the
statistics of the input signal; conversely, the variations of the input will
also be affected by the noise in the processing network\cite{tanasenicola06}. In what follows
below, we assume for simplicity that the spectral-addition rule holds, which means that
the gain, noise and gain-to-noise ratio are independent of the input signal, and that the input does
not need to be specified.

Applying the linearization procedure outlined above may, in general,
qualitatively change the dynamics of the network being considered. However,
previous studies \cite{ziv07, bruggeman09} have shown that the Linear Noise
Approximation provides an accurate description of many systems if the average
copy numbers are of order 10 molecules or more. For the networks considered in
this paper we also compared the power spectra calculated in the linear
approximation with the results of stochastic simulations performed with
Gillespie's algorithm \cite{gillespie76}, and again found good agreement when
protein copy numbers are large (see S1). We therefore expect that the linear
analysis presented in this paper provides an accurate description of the
signalling characteristics of these networks.

\section{Results}

First we study a simple cascade, where ``simple'' means that we consider a
cascade where each component only regulates the activity of the next component
in the cascade; a ``simple'' cascade is thus a cascade without autoregulation,
feedback or feedfoward. We analyse this network in detail such that it can serve
as an instructive example of the method described above. In addition, we will
highlight general features of the results which recur in more complex networks.
We then discuss network motifs including autoregulation and negative feedback
loops, which are commonly observed in biochemical networks.

To understand the effects of autoregulation and feedback we will compare
information transmission in these motifs to a corresponding simple cascade with
the same number of components but without the additional regulation. In order to
perform such a comparison of different motifs on an equal footing we constrain
the average production rate of every component such that these are equal in the
networks under comparison. We argue that from a biological perspective the rate
of protein production is a more significant constraint on network design than
average protein copy number, since the latter only depends on the {\em ratio} of
the synthesis and degradation rate, while it is the absolute synthesis and
degradation rate that determines the cost of having a protein at a particular
copy number. This constraint also enforces that the noise strength at each level
of the cascade $\avgabssq{\eta_i}=2\avg{f^+_i}$ is the same in the motifs being
compared. When comparing two systems with many parameters, equalising production
rates is not a sufficient constraint to uniquely specify all parameter values.
To reduce this potential parameter space we will (unless otherwise stated) hold
constant as many of the network parameters as possible. For brevity we will
only discuss networks in which all regulation occurs via the production
reactions, with linear degradation of each component. However, our results are
qualitatively unchanged if we instead consider regulation via protein
degradation.

We characterise information transmission through these motifs in terms of the
gain, noise and gain-to-noise ratio. Since we assume that the spectral-addition
rule holds \cite{tanasenicola06}, these quantities are intrinsic,
signal-independent properties of the network. We also wish to highlight
differences between the information transmission characteristics of the network,
as determined by the gain-to-noise ratio, and the output power spectrum
$P_{xx}(\omega)$, since this is commonly discussed in studies of signal
transmission. Since $P_{xx}(\omega)$ depends not only on the processing network but
also on the input signal (see Eq. \ref{eqn:Sadd}), we must therefore specify
$P_{ss}(\omega)$; for this purpose we assume, for convenience, that the input
signal $s(t)$ is generated via a Poisson birth-death process as in
Eqn.~\ref{eqn:lin_casc_s} ({\em The simple cascade}).

\subsection{The simple cascade}

Initially we study a simple cascade with a single intermediate component.
Extension of the cascade with more intermediate components is straightforward.
The appropriate reaction scheme is
\begin{subequations} \label{eqn:lin_casc}
\begin{eqnarray}\label{eqn:lin_casc_s}
	\frac{ds}{dt}&=&k_s-\mu_s s+\Gamma\br{t} \\
	\frac{dv}{dt}&=&k_v s-\mu_v v+\eta_v\br{t} \\
	\frac{dx}{dt}&=&k_x v - \mu_x x+\eta_x\br{t}.
\end{eqnarray}\end{subequations}
We reiterate that we assume that there are no cross correlations in the
noise; $\avg{\eta_\alpha(t) \eta_\beta(t^\prime)}=\delta_{\alpha\beta}\delta (t-t^\prime)$ and
$\avg{\Gamma\br{t} \eta_\alpha\br{t'}}=0$. This means that
the reactions are of the type $s \to s+v$ and $v\to v+x$, and not $s\to v$ and $v\to x$,
respectively; put differently, the firing of a reaction does not consume a
molecule of the reactant, and hence does not affect the fluctuations of the
up-stream component \cite{tanasenicola06}. In the {\em Discussion} section, we
will briefly address some of the limitations of this assumption.

Fourier transformation gives
\begin{multline}
	\tilde{X}(\omega)=\overbrace{\frac{k_xk_v \tilde{S}}
		{\br{i\omega+\mu_x}\br{i\omega+\mu_v}}}^{\text{signal}}\\
	+\underbrace{\frac{k_x\eta_v(\omega)}
		{\br{i\omega+\mu_x}\br{i\omega+\mu_v}}
		+\frac{\eta_x(\omega)}{i\omega+\mu_x}
	}_{\text{noise}}.
\end{multline}
As indicated, we can identify the components of the output which are due to the
input $\tilde{S}$ (``signal'') and components which are due to intrinsic noise
in the network. We obtain for the power spectrum of $x$,
\begin{multline}
	P_{xx}(\omega)=\avg{\tilde{X}\tilde{X}^\ast}=
		\overbrace{\frac{k_x^2}{\br{\omega^2+\mu_x^2}}
			\frac{k_v^2}{\br{\omega^2+\mu_v^2}}
		}^{g^2(\omega)}
		\overbrace{\frac{2k_s}{(\omega^2+\mu_s^2)}}^{P_{ss}(\omega)} \\
	+ \underbrace{
			\underbrace{\frac{k_x^2}{\br{\omega^2+\mu_x^2}}
				\frac{2k_v\avg{s}}{\br{\omega^2+\mu_v^2}}
			}_{N_{v\to x}(\omega)}
		+\underbrace{\frac{2k_x\avg{v}}{\br{\omega^2+\mu_x^2}}}_{N_x(\omega)}
		}_{N(\omega)}
	\label{eqn:lin3_PR}
\end{multline}

\begin{figure} [ht]
\begin{tabular}{r l}
\letter{A} & \includegraphics[angle=-90,scale=0.33]{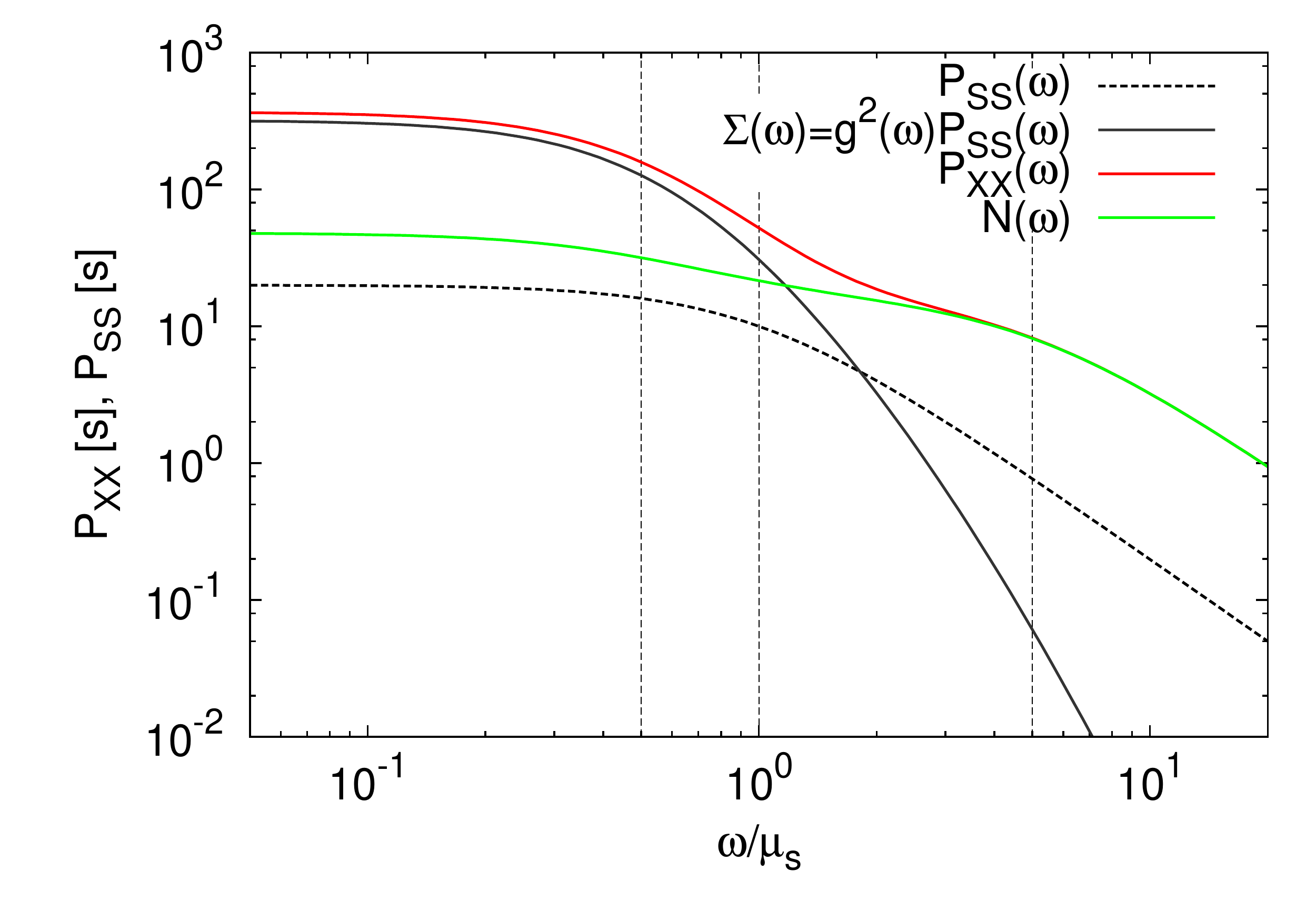}\\
\letter{B} & \includegraphics[angle=-90,scale=0.33]{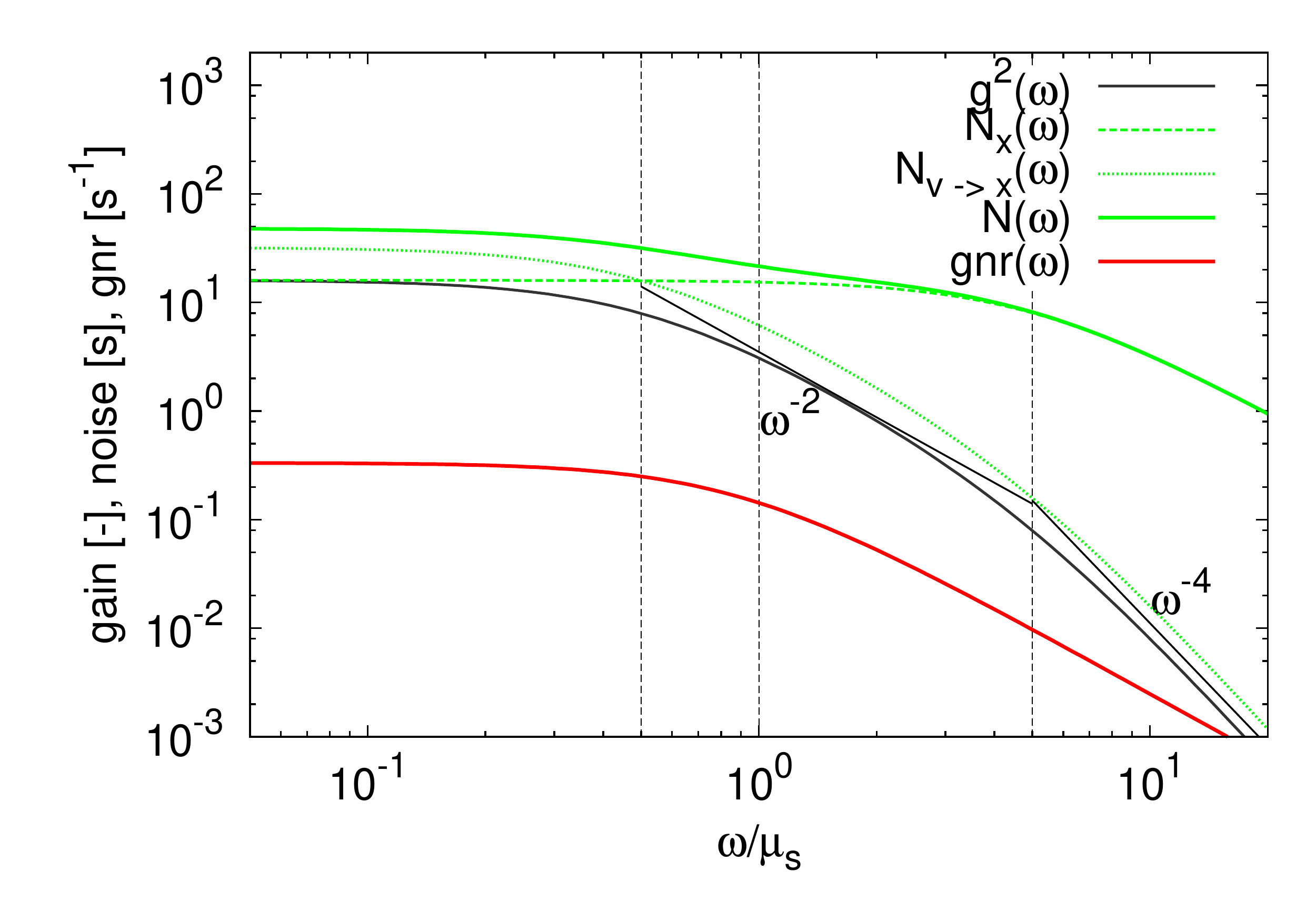}
\end{tabular}
	\caption{ {\bf Typical power spectra for a linear cascade}.
		\textbf{A}: The power spectra of $x$ and $s$, $P_{xx}(\omega)$ and
			$P_{ss}(\omega)$, together with the signal $\Sigma(\omega)$ and noise
			$N(\omega)$ components of the output, for the two-step cascade shown in
			Eqn. \ref{eqn:lin_casc}.
		\textbf{B}: The frequency-dependent gain $g^2(\omega)$, noise $N(\omega)$
			and gain-to-noise ratio (GNR). Thin green lines indicate the two
			noise contributions, $N_{v\to x}(\omega)$ (dotted) and $N_x(\omega)$ (dashed).
		Parameters: $k_s=10$, $k_v=10$, $k_x=1$, $\mu_v=0.5$ and $\mu_x=5$. Vertical
		lines indicate the degradation rates of the three components.\label{fig:pow_spec}}
\end{figure}

Fig.~\ref{fig:pow_spec}A shows the output power spectrum of this network
$P_{xx}(\omega)$ (red), as well as its decomposition into the noise $N(\omega)$ (green) and
the transmitted signal $\Sigma(\omega)=g^2(\omega)P_{ss}(\omega)$ (black solid) (see also Eqn.
\ref{eqn:Sadd}). Simple cascades are characterised by a number of ``knee''
frequencies (vertical dashed), corresponding to the characteristic relaxation rates
of the different components of the network (in this case $\mu_s$, $\mu_v$ and
$\mu_x$). These knee frequencies are the inverse of the response times  of the
components --- e.g. $\mu_v=\tau_v^{-1}$.

In order for the processing network to track variations in the input $s$ on a
time scale $\omega^{-1}$, the network should be able to respond on this time
scale. If any component of the processing network has a longer response time,
this variation in $s$ will be filtered. This filtering can be observed in the
transmitted signal $\Sigma(\omega)$, where at frequencies above the first knee
frequency, $\Sigma(\omega)$ scales with $\omega^{-2}$ and for every
consecutive knee frequency, $\Sigma(\omega)$ decays with an additional factor $\omega^{-2}$
(Fig.~\ref{fig:pow_spec}A).
In effect each level of the cascade acts as a low-pass filter, because the
incoming signal is averaged over the protein response time. Mathematically, the
transmitted signal $\Sigma(\omega)$ can be factored into the input signal
$P_{ss}(\omega)$ (black dashed), and the total gain $g^2(\omega)$ (Fig.~\ref{fig:pow_spec}B, black),
which is independent of the input signal (because we assume that the network does not feed back onto
$s$).
Moreover, the total gain of the network is the product of the gain of each
cascade step: $g^2(\omega)=g^2_{s\to v}(\omega)g^2_{v\to x}(\omega)$; decaying as $\omega^{-4}$ for
$\omega\gg \mu_v, \mu_x$ (Fig.~\ref{fig:pow_spec}B). Consequently, the transmitted signal
$\Sigma(\omega)$ decays as $\omega^{-6}$ for $\omega \gg \mu_s,\mu_v,\mu_x$.

Since we assume that there are no cross-correlations between the different noise
terms, the total noise $N(\omega)$ (green line in Figs.~\ref{fig:pow_spec}A and
\ref{fig:pow_spec}B) is given by the noise-addition rule
\cite{paulsson05, tanasenicola06}, which means that $N(\omega)$ is simply given
by the sum of two independent contributions, $N_{v\to x}(\omega)$ (Fig.~\ref{fig:pow_spec}B, green
dotted) and $N_x(\omega)$ (green dashed) (see Eqn. \ref{eqn:lin3_PR}). Here, $N_{x}(\omega)$ is the
noise in the concentration of $x$ that arises from the intrinsic stochasticity in the
production and decay events of $x$; $N_x(\omega)$ would be the total variance in
the concentration of $x$ if $v$, the input for $x$, would not vary over time.
However, the upstream component $v$ does vary in time, not only because it
is driven by variations in the input $s$, but also because it fluctuates
spontaneously due to the noise in its synthesis and decay events. This noise is
propagated to $x$. Its contribution to the total noise power of $x$ is $N_{v\to
x}(\omega)$, which is given by the noise in $v$, $N_{v}(\omega)$, multiplied by
how much this noise is amplified at the level of $x$, given by $g^2_{v\to
x}(\omega)$: $N_{v\to x}(\omega)=g^2_{v\to x}(\omega)N_v(\omega)$, where
$g^2_{v\to x} = k_x^2 / (\omega^2 + \mu_x^2)$. The ``extrinsic'' contribution to
the noise in $x$, $N_{v\to x}$, decays as $\omega^{-4}$ since the noise in $v$,
decaying as $\omega^{-2}$, is filtered by the finite lifetime of the protein
$x$. The ``intrinsic'' contribution, $N_{x}(\omega)$, decays as $\omega^{-2}$,
meaning that for $\omega \gg \mu_v,\mu_x$, $N(\omega) \approx N_x(\omega)$.
Hence, while the transmitted signal $\Sigma(\omega)$ decays as $\omega^{-6}$ for
$\omega \gg \mu_s,\mu_v,\mu_x$, the noise $N(\omega)$ decays as $\omega^{-2}$
(Fig.~\ref{fig:pow_spec}B, green solid).
As a result, for frequencies $\omega \gg \mu_s,\mu_v,\mu_x$, the transmitted
signal $\Sigma(\omega)$ is completely obscured by the noise and the output
$P_{xx}(\omega)$ is simply given by the noise $N(\omega)$ (Fig.
\ref{fig:pow_spec}A).

Finally, the gain-to-noise ratio (Fig.~\ref{fig:pow_spec}B, red) is
\begin{equation} \label{eqn:gnr_2step}
	\frac{g^2(\omega)}{N(\omega)}=\frac{k_vk_x\mu_v}
		{2\avg{s}\left[\omega^2+\mu_v^2+\mu_vk_x\right]}.
\end{equation}
This expression shows that the simple cascade effectively acts as a low-pass
filter for information, meaning that it cannot reliably respond to signals that
vary (much) faster than a characteristic cut-off frequency
$\omega^2_{c}=\mu_v\br{\mu_v+k_x}$. We note that the gain-to-noise ratio is
independent of $\mu_x$, since both the gain and the noise have the same
functional dependence on $\mu_x$. This is a general feature of the biochemical
networks we will study: degradation of the output species occurs independently
of the upstream components, and therefore provides no additional information
about the input \cite{tostevin09}.

\subsection{Autoregulation}

In this section we consider direct feedback of a component onto its own
production, as indicated in Fig.~\ref{fig:autoreg_res}A and
\ref{fig:autoreg_int}A. Autoregulation is one of the most common forms of
regulation in signalling networks. It is well known that negative autoregulation
speeds up the response time of components, which can also change the response
time of the complete signalling cascade \cite{heinrich02}. Positive
autoregulation slows down the response time and can lead to bistability
\cite{heinrich02, alon}.

Autoregulation alters only the diagonal entries of the Jacobian matrix
(Fig.~\ref{fig:mat_jac}). This means that the characteristic timescale for
dissipation of small fluctuations --- the response time --- changes, which is as
expected. For the steady state of the system to be stable we require that the
diagonal of the Jacobian has only negative terms. Thus autoregulation cannot
qualitatively change the form of the output power spectrum $P_{xx}(\omega)$. In
fact, once linearized, the dynamics of a network with autoregulation is
equivalent to that of a simple cascade with a different degradation rate. In
terms of information transmission, however, this is not always true, as we shall
see below.

\subsubsection{Autoregulation at the response $x$ does not affect information transmission}

\begin{figure*}[ht]
\fcolorbox{white}{LightBlue} {
	\begin{minipage}{1.0\textwidth}
		\letter{A}
		\hspace{0.5cm}
		\begin{minipage}{0.90\colwidth}
			\includegraphics[scale=1.0]{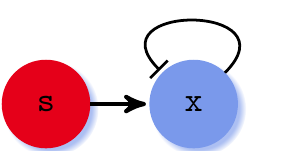}
		\end{minipage}
		\hspace{-0.5cm}
		\letter{B}
		\hspace{0.5cm}
		\begin{minipage}{0.9\colwidth}
		\flushleft
			\begin{equation}\label{eqn:autoreg_res_net}
				\diff{x}{t}=f\br{x}s -\mu_x x +\eta_x,
			\end{equation}
			\begin{equation*}
			f\br{x}=\frac{\nu \beta}{K+x}
 				\begin{cases}
 					\beta=K, &	\text{negative regulation} \\
 					\beta=x, & \text{positive regulation}
				\end{cases}
			\end{equation*}
			\begin{equation*}
				J_{xx}=-\mu_x+\avg{s}\sqbr{\pdiff{f\br{x}}{x}}_{\rm s.s.}
			\end{equation*}
		\end{minipage}
		\vspace{0.5cm}
		\newline
		\flushleft
		\letter{C}
		\begin{minipage}{0.90\colwidth}
			\begin{subequations}\label{eqn:autoreg_res_eqns}
			\begin{equation}
			g^2=\frac{J_{xs}^2}{\omega^2+J_{xx}^2}
			\end{equation}
			\begin{equation}
			N=\frac{\avgabssq{\eta_x}}{\omega^2+J_{xx}^2}
			\end{equation}
			\begin{equation}
			\frac{g^2}{N}=\frac{J_{xs}^2}{\avgabssq{\eta_x}}
			\end{equation}
			\end{subequations}
		\end{minipage}
	\end{minipage}
}

	\caption{{\bf Autoregulation of the output component}.
		\textbf{A}: Schematic representation of the negative autoregulation motif,
			where $s$ is the input signal and $x$ the output signal, which negatively
			regulates its own production.
		\textbf{B}: The Langevin equations of the network.
		\textbf{C}: The characteristic equations for the gain, noise and
			gain-to-noise ratio (see also section {\em
Autoregulation}).\label{fig:autoreg_res}
	}
\end{figure*}

We first consider autoregulation by the network output $x$ on its own
production, as depicted in Fig.~\ref{fig:autoreg_res}A. For this motif the
relaxation time of $x$ is given by
$\tau_x=-J_{xx}^{-1}=\sqbr{\mu_x-\avg{\pdiff{}{x}f(x)s}}^{-1}$, where $f(x)$
describes the effect of the feedback of $x$ onto its own production (see Eqn.
\ref{eqn:autoreg_res_net} in Fig.~\ref{fig:autoreg_res}B) For negative regulation
$\abs{J_{xx}}>\mu_x$, while
for positive regulation $\abs{J_{xx}}<\mu_x$. Negative (positive) regulation
therefore reduces (increases) the response time of $x$ to changes in $s$,
compared to the equivalent simple cascade network for which $f(x)=$constant.
In the output power spectrum $P_{xx}(\omega)$ this change in timescale appears
as a shift in the knee frequency corresponding to $\tau_x^{-1}$. A corresponding
change can also be seen in both the gain and noise (see Eqn.~\ref{eqn:autoreg_res_eqns} in
Fig.~\ref{fig:autoreg_res}C).

However, despite these changes in the response time, we find that the
gain-to-noise ratio for an autoregulatory network
(Eqn.~\ref{eqn:autoreg_res_eqns}c in Fig.~\ref{fig:autoreg_res}B) is
identical to the gain-to-noise ratio for a simple (two-component)
cascade. The effect of changing $J_{xx}$ on the noise and gain is
identical (Eqn.~\ref{eqn:autoreg_res_eqns}a,b in Fig.~\ref{fig:autoreg_res}B) and therefore cancels
in the gain-to-noise ratio (as we also saw previously for the effect of
$\mu_x$ in the simple cascade, Eqn.~\ref{eqn:gnr_2step}). The
autoregulation by $x$ of its own production alters the timing of
production events. However, our constraint of equal average production
means that the mean rate of this process in the two cascades is the
same. Moreover, in the linearized regime the production of $x$ is an
identical Poissonian process in both simple and autoregulated
cascades. Hence, to the extent that the system can be linearized,
autoregulation at the output of a network does not affect information
transmission. It is conceivable that non-linear effects cause
autoregulation of the output component to affect information
transmission, but a comparison of our analytical results discussed
here with results of Gillespie simulations of the full system, suggest
that the linearization approximation is surprisingly accurate (see
also S1).

\subsubsection{Positive autoregulation within the cascade increases the
gain-to-noise ratio}

\begin{figure*}[!ht]
\fcolorbox{white}{LightBlue} {
	\begin{minipage}{0.99\textwidth}
	\flushleft
		\letter{A}
		\hspace{0.5cm}
		\begin{minipage}[t]{0.90\colwidth}
			\includegraphics[scale=1.0]{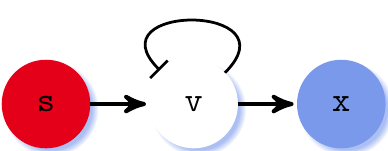}
		\end{minipage}
		\hspace{-0.7cm}
		\letter{B}
		\hspace{0.5cm}
		\begin{minipage}{0.90\colwidth}
			\begin{subequations}
			\label{eqn:autoreg_int_net}
			\begin{eqnarray}
				\diff{v}{t}&=&f(v)s -\mu_v v +\eta_v \\
				\diff{x}{t}&=&\beta v -\mu_x x +\eta_x.
			\end{eqnarray}
			\begin{equation*}
			f\br{v}=\frac{\nu \beta}{K+v}
 				\begin{cases}
 					\beta=K, &	\text{negative regulation} \\
 					\beta=v, & \text{positive regulation}
				\end{cases}
			\end{equation*}
			\end{subequations}
			and
 			\begin{equation}
				J_{vv}=-\mu_v+\avg{s}\sqbr{\pdiff{f\br{v}}{v}}_{\rm s.s.}
			\end{equation}
		\end{minipage}
			\vspace{0.5cm}
			\newline
			\flushleft
		\letter{C}
		\begin{minipage}[t]{0.90\colwidth}
			\begin{subequations}\label{eqn:autoreg_int_eqns}
			\begin{equation}\label{eqn:autoreg_int_eqns_gain}
				g^2=\frac{\br{J_{xv}J_{vs}}^2}{\br{\omega^2+\mu_x^2}
				\br{\omega^2+J_{vv}^2}}
			\end{equation}
			\begin{equation}\label{eqn:autoreg_int_eqns_noise}
				N = \frac{J_{xv}^2\avgabssq{\eta_v}+\br{\omega^2+J_{vv}^2}
					\avgabssq{\eta_x}}{\br{\omega^2+\mu_x^2}
					\br{\omega^2+J_{vv}^2}}
			\end{equation}
			\begin{equation}
				\frac{g^2}{\ds N}=\frac{J_{xv}J_{vs}\mu_v}
					{\ds 2\avg{s}\sqbr{\omega^2+J_{vv}^2+J_{xv}\mu_v}}
			\end{equation}
		\end{subequations}
		\end{minipage}
		\letter{D}
		\hspace{0.5cm}
		\begin{minipage}[t]{0.90\colwidth}
			\includegraphics[angle=-90,scale=0.28]{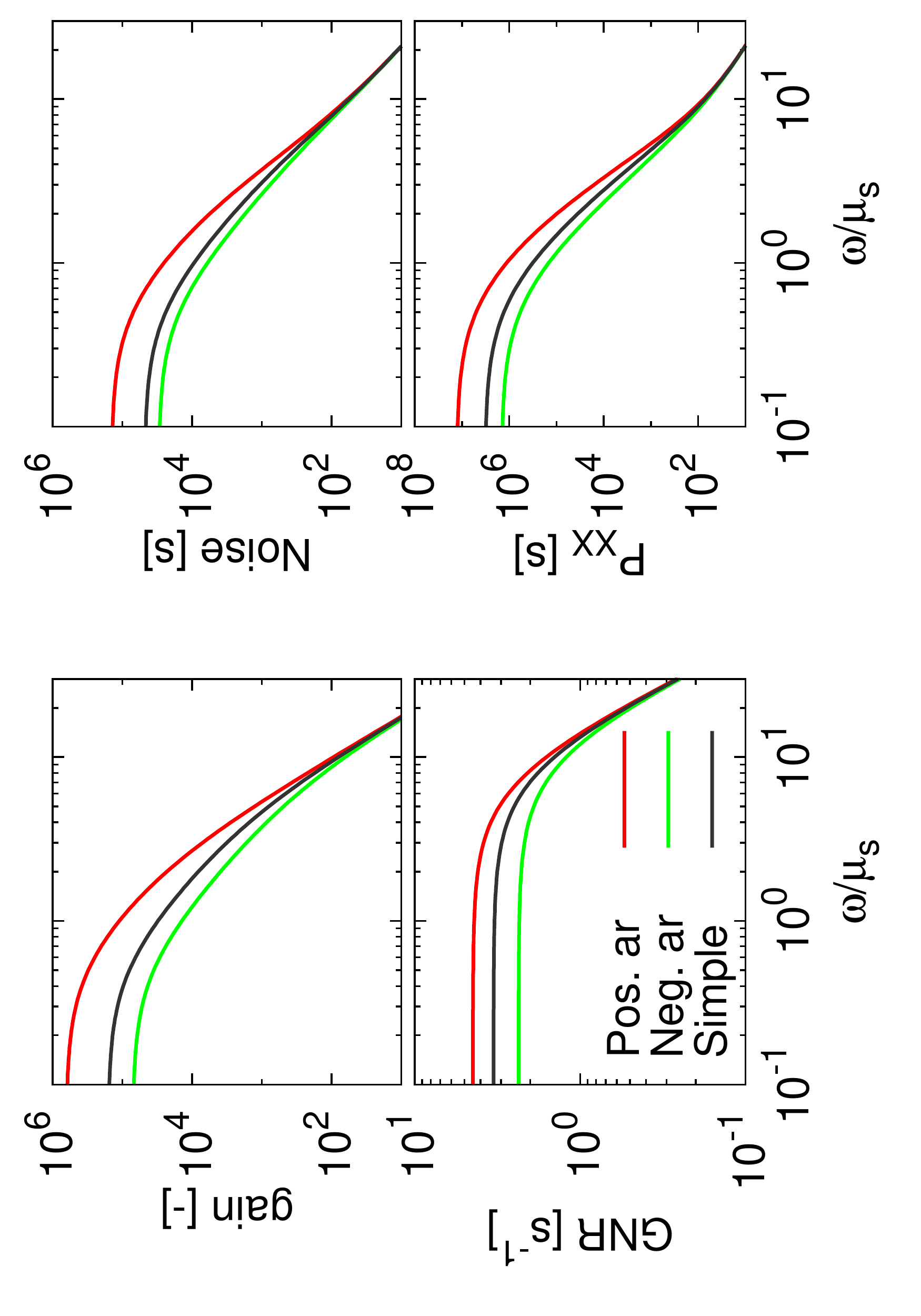}
		\end{minipage}
		\newline
		\letter{E}
		\begin{minipage}[t]{1.0\colwidth}
			\includegraphics[angle=-90,scale=0.28]{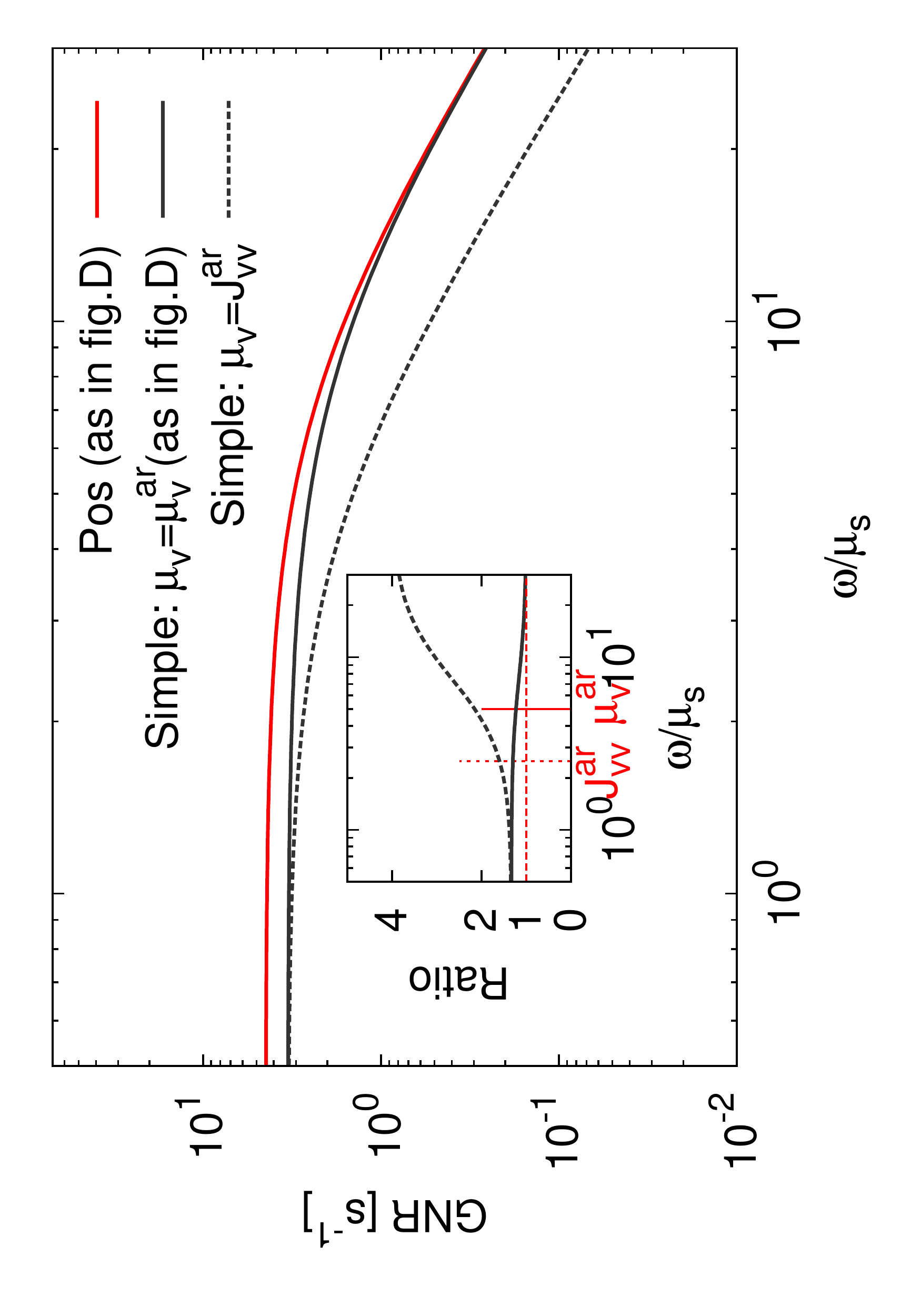}
		\end{minipage}
	\end{minipage}
}

	\caption{{\bf A two-step cascade with autoregulation of the
			intermediate component}.
		\textbf{A}: Cartoon of the negative autoregulation motif, where the
			intermediate component $v$ negatively regulates its own production.
		\textbf{B}: The Langevin equations describing the network.
		\textbf{C}: The characteristic equations for the gain $g^2(\omega)$, noise
			$N(\omega)$ and gain-to-noise ratio.
		\textbf{D}: The gain, noise, gain-to-noise ratio (GNR) and output power
			spectrum $P_{xx} (\omega)$ plotted as a function of frequency for three
			different cascades: simple (black), positive autoregulation (red) and
			negative autoregulation (green). Negative autoregulation reduces the gain,
			noise and gain-to-noise ratio. For positive autoregulation the opposite
			holds. Positive autoregulation has a smaller knee frequency in the
			gain-to-noise ratio than negative autoregulation (see also section
			{\em Autoregulation}). Parameters: $k_s=10$, $k_v=100$, $k_x=10$,
			$\mu_v=5$, $\mu_x=0.5$, $K=\avg{v}$ and $\nu^a=200$, $\nu^r=200$.
		\textbf{E}: The gain-to-noise ratio for a cascade with postive autoregulation (red
			line) and two simple cascades (black). Solid: as in $\textbf{D}$, the
			degradation rate of the simple cascade equals that of the cascade with
			positive autoregulation, $\mu_v=\mu^{ar}_v$. Hence $\abs{J_{vv}}$ is
			smaller in the cascade with auto-activation, and the gain-to-noise ratio
			is larger at low frequencies. Dotted: In the simple cascade we take
			$\mu_v=J^{\rm ar}_{vv}$, and instead increase the production rate
			$J_{xv}$. This decreases the gain-to-noise ratio of the simple cascade with respect to the
autoregulated cascade over the full frequency spectrum. Inset: the ratio of the gain-to-noise
ratio of the cascade with	positive autoregulation to that of the simple cascade; solid:
$\mu_v=\mu^{ar}_v$,	dotted $\mu_v=J^{ar}_{vv}$. The dashed red vertical line indicates
			$J^{ar}_{vv}$, the vertical solid red line $\mu^{ar}_v$, which shows the
			shift in frequency dependence. Parameters: $k_s=10$, $k_v=100$, $k_x=10$,
			$\mu_v=5$, $\mu_x=0.5$, $K=\avg{v}$ and $\nu=200$.\label{fig:autoreg_int}
	}
\end{figure*}

In a cascade with autoregulation by an intermediate component the
story is different (Fig.~\ref{fig:autoreg_int}A and Eqn.~\ref{eqn:autoreg_int_net} in
Fig.~\ref{fig:autoreg_int}B). First,
we reiterate that since we compare the simple cascade and the cascade with autoregulation on the
basis of equal average production and degradation rates, the noise strengths $\avg{\eta_x^2}$ and
$\avg{\eta_v^2}$ are the same for both cascades.
However, as noted above the effective relaxation timescale of component $v$, $\tau_v=-J^{-1}_{vv}$
(Eqn.~\ref{eqn:lin_sys}), decreases with negative autoregulation and increases with positive
autoregulation. This again leads to a reduction (increase) in both the
gain (Fig.~\ref{fig:autoreg_int}D, top left) and the noise (Fig.~\ref{fig:autoreg_int}D, top
right)
of the network for negative (positive) autoregulation, as has been reported previously
\cite{thattai01, simpson03}. However, unlike the case of autoregulation of the output $x$, the
gain-to-noise ratio (Fig.~\ref{fig:autoreg_int}D, bottom left) can change as a result.

Negative autoregulation (Fig.~\ref{fig:autoreg_int}D, green)  leads to a decrease in the
response time compared to a simple cascade, corresponding to an increase in $\abs{J_{vv}}$. This
leads to a decrease in the gain of the autoregulated component $g^2_{s\to v}(\omega)=
J_{vs}^2/(\omega^2+J_{vv}^2)$ at frequencies $\omega<\abs{J_{vv}}$. Negative autoregulation
therefore tends to suppress slowly varying signals relative to the simple cascade. Noise which is
introduced upstream of or at the autoregulated component is filtered by the feedback-modified gain
in exactly the same way as the signal, whereas noise introduced downstream of $v$ is unaffected.
Hence negative autoregulation reduces both the total gain of the network, which is the product of
the individual reaction gains $g^2(\omega)= g^2_{s\to v}(\omega)g^2_{v\to x}(\omega)$, and the noise
transmitted from $v$ to $x$, $N_{v\to x}(\omega)=g^2_{v\to x}N_v(\omega)$, relative to the simple
cascade. However, noise in the production and degradation of $x$ is unchanged relative to the simple
cascade. Since the total noise (Eqn.~\ref{eqn:autoreg_int_eqns}b in
Fig.~\ref{fig:autoreg_int}C) is the sum of independent
noise contributions, $N(\omega)=N_x(\omega)+N_{v\to x}(\omega)$, the total noise decreases by a
smaller factor than the gain, and the gain-to-noise ratio decreases compared with the simple
cascade.

Conversely, positive autoregulation (Fig.~\ref{fig:autoreg_int}D, red) increases the relaxation
time of $v$, which increases $g^2_{s\to v}(\omega)$ at frequencies $\omega<\abs{J_{vv}}$. We can
therefore see that positive autoregulation amplifies slowly-varying signals. This leads to an
increase in the network gain and the noise that is propagated from $v$ to $x$. However, since the
noise that is introduced at $x$ is unchanged, positive autoregulation at $v$ increases the
gain-to-noise ratio compared to the simple cascade. Fig.~\ref{fig:autoreg_int}D shows the
comparison between a simple cascade and cascades with positive (red) and negative (green)
autoregulation.
Hornung and Barkai previously studied transmission of a \textit{constant} signal with additive noise
through a deterministic (noiseless) network \cite{hornung08}, and found that positive autoregulation
can increase the signal-to-noise ratio. Our results for time-varying signals with intrinsic network
noise parallel their results.

Given a network with autoregulation, our constraint of equal production of each network component
does not define a unique ``equivalent'' simple cascade.  That is, different parameter combinations
can be chosen for a simple cascade which satisfy the production constraint. The results in the
preceding discussion correspond to one such parameter choice. Specifically, we choose the
production rate of $v$ in the simple cascade (Eqn.~\ref{eqn:lin_casc}) to be $k_v=\avg{f(v)}$, while
taking the same value for $\mu_v$ in both networks. A consequence of this choice is that the
relaxation time $\tau_v$ changes between the two cascades, as discussed above. One can equally well
construct a simple cascade for which the diagonal entries of the Jacobian, $J_{\alpha\alpha}$, are
equal to those of the autoregulated cascade, so as to hold constant the relaxation time of each
component between the two cascades. This is achieved by setting the spontaneous degradation rate for
$v$ in Eqn.~\ref{eqn:lin_casc} to be $\mu_v^{\rm new}=\mu_v-\avg{\pdiff{}{v}f(v)s}$. By choosing
this new rate, the average protein number $\avg{v}$ changes in the simple cascade, and as a result
also the average production of $x$. To restore equal production of $x$ we thus also require
a rescaling of the kinetic production rate $k_x^{\rm new}=k_x\mu_v^{\rm new}/\mu_v$ in the simple
cascade (Eqn. \ref{eqn:lin_casc}). Thus, in this comparison, the diagonal entries of the Jacobian
matrices of the autoregulated and simple cascade are the same, while the off-diagonal entry
$J_{xv}=k_x$ differs between the two.

Compared to a cascade with positive autoregulation, this new kinetic production rate in the simple
cascade is smaller ($k_x^{\text{new}}<k_x$). The reduction in $J_{xv}$ leads to a uniform
decrease in $g^2_{v\to x}(\omega)$ at all frequencies. As described above, this affects the signal
and also the propagated noise $N_{v\to x}(\omega)$ equally, but not the intrinsic noise at $x$,
$N_x(\omega)$. Thus, compared to a cascade with positive autoregulation, the gain-to-noise ratio is
reduced at all frequencies in the simple cascade, as can be seen in Fig.~\ref{fig:autoreg_int}E
(black dotted). Interestingly, the decrease in the gain-to-noise ratio is most pronounced at high
frequencies. This is because the propagated noise $N_{v\to x}(\omega)$ only has a significant
contribution at frequencies $\omega<\mu_v^{\rm new}$; at higher frequencies the total noise is
dominated by $N_x(\omega)$, as discussed in section: {\em The simple cascade}. Thus at these higher
frequencies, the gain is reduced relative to the positively-autoregulated cascade, but the noise is
not, and so the change in the gain-to-noise is largest. For networks with negative autoregulation,
the converse applies: the gain-to-noise ratio is higher in the simple cascade at all frequencies,
but by a larger factor for $\omega<\mu^{\rm new}_v$. Hence, the effect of positive or negative
autoregulation is qualitatively the same in both parameterizations.

More generally, even if we relax the production constraint on each component, and instead require
only the total production in the two cascades to be the same (i.e., $\avg{f^+_v}+\avg{f^+_x}={\rm
constant}$), we see the similar qualitative behaviour for the gain-to-noise ratio
(see Eqn.~S1-26-S1-28). Positively-autoregulated cascades have a larger gain-to-noise ratio than a
simple
cascade of the same length, while for a cascade with negative autoregulation the gain-to-noise ratio
is smaller. For longer cascades drawing such general conclusions is more difficult. However, if the
majority of parameters are kept the same between the simple and autoregulated cascades, as in the
cases discussed in detail above, then we again find that positive autoregulation increases and
negative autoregulation decreases the gain-to-noise ratio. Furthermore, given a specific simple
cascade one can always add positive autoregulation to the network in such a way as to achieve a
larger gain-to-noise ratio while maintaining the same total production cost.

We have here considered only autoregulation via the production of the intermediate $v$. However, for
autoregulation via the degradation of $v$ we observe similar results for the gain-to-noise ratio: if
$v$ suppresses its own degradation, the decrease in the effective turn-over rate leads to a
reduction of the noise strength $N_{v\to x}(\omega)$, increasing the gain-to-noise ratio; when $v$
enhances its own degradation rate the transmitted noise is increased, reducing signalling fidelity.

\subsection{Feedback}

Feedback, both positive and negative, corresponds to the upper-triangular part in the Jacobian of
the linearized system (see Fig~\ref{fig:mat_jac}).  It is known that negative feedback allows for
adaptation as, for example, in the {\em E. coli} chemotaxis pathway \cite{macnab72, segall83, tu08}.
Feedback can also shift noise to higher frequencies \cite{simpson03}. We will again consider
separately the two cases of feedback by the output $x$ onto an upstream component and feedback by an
intermediate component onto a component higher up the cascade.

\subsubsection{Feedback from $x$ does not affect information transmission}

\begin{figure*}[!ht]
	\fcolorbox{white}{LightBlue} {
	\begin{minipage}{0.99\textwidth}
	\flushleft
		\letter{A}
		\hspace{0.5cm}
		\begin{minipage}[b]{0.90\colwidth}
			\includegraphics[scale=1.0]{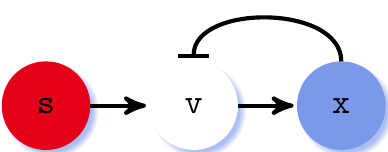}
		\end{minipage}
		\hspace{-0.5cm}
		\letter{B}
		\hspace{1.0cm}
		\begin{minipage}{0.90\colwidth}
		\raggedright
			\begin{subequations}
			\label{eqn:fb_res_net}
			\begin{eqnarray}
				\diff{v}{t}&=&f\br{x}s-\mu_{v}v+\eta_v \\
				\diff{x}{t}&=&\beta v-\mu_{x}x+\eta_x,
			\end{eqnarray}\end{subequations}
			where
			\begin{equation*} f\br{x}=\frac{\nu C^n}{K^n+x^n}
				\begin{cases}
				C=x,& \text{positive feedback}\\
				C=K,& \text{negative feedback}\\
				\end{cases}
			\end{equation*}
		\end{minipage}
			\vspace{0.5cm}
			\newline
		\letter{C}
		\hspace{-0.5cm}
		\begin{minipage}[t]{0.90\colwidth}
		\raggedright
		\begin{subequations}\label{eqn:fb_res}
		\begin{equation}\label{eqn:fb_res_gain}
		g^2=
\frac{\br{J_{vs}J_{xv}}^2}{\br{\omega^2+\mu_v^2}\br{\omega^2+\mu_x^2}
+F(\omega)}
		\end{equation}
		\begin{equation}\label{eqn:fb_res_noise}
		N =\frac{\beta^2\avgabssq{\eta_v}+(\omega^2+\mu_v^2)\avgabssq{\eta_x}}
{\br{\omega^2+\mu_v^2}\br{\omega^2+\mu_x^2} +
F(\omega)}
		\end{equation}
		\begin{equation} \label{eqn:fb_res_gnr}
		\frac{g^2}{N}=\frac{(J_{vs}J_{xv})^2} {
\beta^2\avgabssq{\eta_v}+(\omega^2+\mu_v^2)\avgabssq{\eta_x}},
		\end{equation}
		\begin{equation}\label{eqn:fb_res_pxx}
P_{xx}(\omega)=\frac{J^2_{vs}J^2_{xv}P_{ss}(\omega)+J_{xv}
^2\avgabssq {
\eta_v}+\br{\omega^2+\mu_v^2}\avgabssq{\eta_x}} {\br{\omega^2+\mu_v^2}\br{
\omega^2+\mu_x
^2}+F(\omega)}
\end{equation}
		\end{subequations}
		with
		\begin{equation}
		F(\omega)=J_{vx}J_{xv}\sqbr{J_{vx}J_{xv}+2\br{\omega^2-\mu_x\mu_v}}
		\end{equation}
		\end{minipage}
		\hspace{0.6cm}
		\letter{D}
		\hspace{0.5cm}
		\begin{minipage}[t]{0.90\colwidth}
			\includegraphics[angle=-90,scale=0.28]{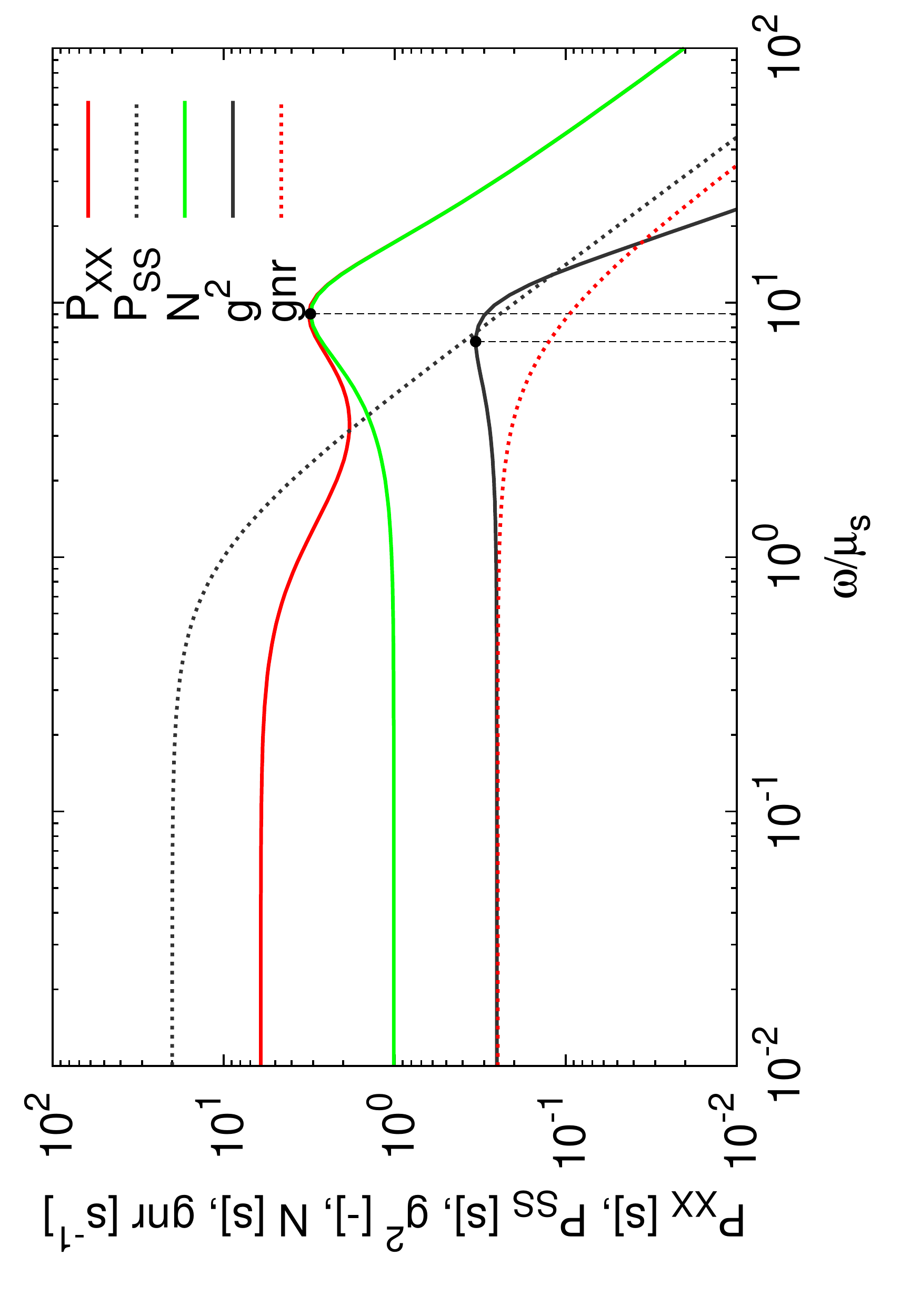}
		\end{minipage}
	\end{minipage}
}

	\caption{ {\bf Feedback from the output signal $x$ to an upstream
			component, discussed in section {\em Feedback }}.
		\textbf{A}: Cartoon of the negative feedback motif, where the output
			signal $x$ negatively regulates $v$.
		\textbf{B}: The Langevin equations describing the network.
		\textbf{C}: The characteristic equations: gain $g^2(\omega)$, noise
			$N(\omega)$ and gain-to-noise ratio $g^2(\omega)/N(\omega)$.
		\textbf{D}: Power spectra of the output, $P_{xx}(\omega)$, input,
			$P_{ss}(\omega)$, gain $g^2(\omega)$ and noise $N(\omega)$.
			$P_{xx}(\omega)$, $g^2(\omega)$ and $N(\omega)$ all exhibit a peak due to
			the negative feedback, while the gain-to-noise ratio is monotonically
			decreasing. The black dots indicate the peaks of the gain and noise, which
			occur at different frequencies (see section {\em Feedback}).
			Parameters: $k_s=10$, $K=0.2\avg{x}$, $\nu=1260$, $k_x=5$, $\mu_v=5$,
			$\mu_x=5$ and $n=3$.\label{fig:fb_res}
	}
\end{figure*}

For negative feedback from $x$ to $v$ (Fig.~\ref{fig:fb_res}A and Eqn.~\ref{eqn:fb_res_net} in Fig.~
\ref{fig:fb_res}B),
the power spectrum of the response $P_{xx}(\omega)$ (Fig.~\ref{fig:fb_res}D, red solid) can have a
resonance peak while none is present in the input signal (black dotted). Surprisingly, this peak
does not correspond to an increase in information transmission capabilities at the peak frequency
($\omega_{\rm peak}$), since no peak is present in the gain-to-noise ratio (Fig.~\ref{fig:fb_res}D,
red dashed). For positive feedback, no peak is present in either $P_{xx}(\omega)$ or the
gain-to-noise ratio.

For a system with negative feedback from $x$ to $v$ the gain and noise both show a peak, but these
can occur at different frequencies. We consider first the frequency dependence of the gain. At low
frequencies the negative feedback leads to destructive interference at $v$ between the input
signal $\tilde{S}(\omega)$ and the signal that is fed back, $\tilde{X}(\omega)$. On the other hand,
at high frequencies these two signals are exactly out of phase, and hence the interference becomes
constructive (since the feedback combines the two signals negatively). However, at frequencies
$\omega\gg\mu_v,\mu_x$ the {\em amplitude} of the fed-back signal decreases, due to averaging over
the lifetimes of $v$ and $x$; hence, even though the two signals interfere constructively, the
significance of this interference decreases. Together, these three effects lead to a maximum in the
gain. This maximum occurs at 
\begin{equation} \label{eqn:omega_res}
	\omega_{\rm res}^2=-\frac{1}{2}\left[\mu_x^2+\mu_v^2+2J_{vx}J_{xv}\right],
\end{equation}
which depends on the relaxation rates $\mu_x$, $\mu_v$ and the coupling
(feedback) loop between $v$ and $x$, $J_{vx}J_{xv}$. This timescale corresponds
to the imaginary part of the eigenvalues of the Jacobian (see Eqn.~S1-39).

The frequency of the peak in the noise depends on the relative strengths of the
two noise sources, $\eta_v$ and $\eta_x$. The two noise terms are propagated
differently through the network, because $\eta_x$ originates at the regulator of
the feedback loop, while $\eta_v$ originates at the regulated component. We
consider two limiting cases. If the total noise $N(\omega)$
(Eqn.~\ref{eqn:fb_res}b in Fig.~\ref{fig:fb_res}C) is dominated by the transmitted noise,
$N_{v\to x}(\omega)$, both the signal $\Sigma(\omega)$ and the dominant source of noise
originate upstream of the feedback loop. Effectively, therefore, the feedback
affects both the gain and noise of the network similarly. As a result the peak
frequencies of both the noise and the gain are the same. On the other hand, when
the total noise is dominated by $N_x(\omega)$, which is located downstream of
the regulated component $v$, the feedback loop affects the signal and noise
differently. As a result, the noise that is fed back has a different frequency
profile than the signal, such that the peaks in the gain and the noise occur at
different frequencies (Fig.~\ref{fig:fb_res}D, black circles).

One might therefore expect that when $N_x(\omega)\gg N_{v\to x}(\omega)$ a peak
in the gain-to-noise ratio is possible. However, an inspection of the
expressions for the gain, Eqn.~\ref{eqn:fb_res}a, and the noise,
Eqn.~\ref{eqn:fb_res}b (both in Fig.~\ref{fig:fb_res}C), shows that they have the same
denominator, such that the gain-to-noise ratio is a monotonically decreasing function of frequency
(Eqn.~\ref{eqn:fb_res}c in Fig.~\ref{fig:fb_res}C). The effect of the negative feedback is
cancelled.
Ultimately, this is due to the fact that the noise in the output $x$ goes back
into the feedback loop, such that the peaks in the gain and the noise cannot be
controlled separately; in the next section, we show how this can be done.
Furthermore, we note that the gain-to-noise ratio is again identical to a simple
three-component cascade, as we also saw in the case of autoregulation of $x$. We
conclude that feedback from $x$ onto the cascade also has no effect on
information transmission through the network.

This network (Eqn.\ref{eqn:fb_res} in Fig.~\ref{fig:fb_res}B) also highlights the idea that the
power spectrum of
the output $P_{xx}(\omega)$ may not be indicative of the information that is transmitted at
different frequencies. We see in Fig.~\ref{fig:fb_res}D that due to the negative
feedback $P_{xx}(\omega)$ can have a peak at non-zero frequencies, even if none
is present in the input signal. However, this peak does not correspond to the
frequency at which the signal is transmitted most reliably. Instead, we can see
that the peak is simply due to resonant amplification of both the signal and the
noise at the characteristic frequency of the negative feedback loop.

It has been suggested \cite{locasale08} that a system where a negative feedback
loop acts {\em on} the response component can have a large peak in the gain,
such that signals on specific timescales can be selected for. If we take in Fig.~\ref{fig:fb_res}A
not $x$ but $v$ to be the output of the network, we obtain
\begin{equation}
	\frac{g^2(\omega)}{N(\omega)}=
		\frac{J_{vs}^2\br{\omega^2+\mu_x^2}}
			{J_{vx}^2\avgabssq{\eta_x} +\br{\omega^2+\mu_x^2}\avgabssq{\eta_v}}
\end{equation}
We observe that the gain-to-noise ratio is a monotonically increasing function
of frequency and does not show a peak at any specific frequencies. Furthermore
we note that as $\omega\to\infty$ the gain-to-noise ratio becomes equal to the
gain-to-noise ratio for the one-step simple cascade ($J_{vs}/2\avg{s}$), since
for large $\omega$ the noise from the downstream component is averaged out. Thus
this network motif has a higher gain-to-noise at all frequencies than the
cascade with $x$ as the output. However, the information transmitted at low
frequencies is less than if $x$ were not present. Following the information
processing inequality, the amount of information about $s$ which is encoded in
the dynamics of $v$ is always larger than the corresponding information in $x$.
By feeding back $x$ to $v$ we thus do not add more information to the signal,
but essentially add an extra source of noise to the pathway from $s$ to $v$. The
strength of this noise is highest at frequencies $\omega<\mu_x$, and hence the
effect of the feedback is to obscure the signal at these frequencies. As a
result this motif acts as a high-pass filter for information.

\subsubsection{Negative feedback within a cascade can lead to a peak in the
gain-to-noise ratio}

\begin{figure*}[!ht]
\fcolorbox{white}{LightBlue} {
	\begin{minipage}{0.99\textwidth}
	\flushleft
	\letter{A}
		\hspace{0.5cm}
		\begin{minipage}{0.90\colwidth}
			\includegraphics[scale=1.0]{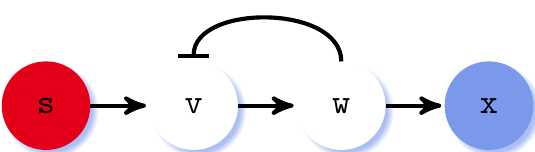}
		\end{minipage}
		\hspace{0.3cm}
		\letter{B}
		\begin{minipage}{0.90\colwidth}
		\begin{subequations}
			\label{eqn:fb_int_net}
			\begin{eqnarray}
			\diff{v}{t}&=&f\br{w}s-\mu_{v}v+\eta_{v}\label{eqn:fb_int_net_1}\\
			\diff{w}{t}&=&\beta v -\mu_{w}w+\eta_{w}\\
			\diff{x}{t}&=&\gamma w -\mu_x x+\eta_x \label{eqn:fb_int_net_3},
			\end{eqnarray}\end{subequations}
				\begin{equation*} f\br{w}=\frac{\nu C^n}{K^n+w^n}
				\begin{cases}
				C=w,& \text{positive feedback}\\
				C=K,& \text{negative feedback}\\
				\end{cases}
			\end{equation*}
		\end{minipage}
		\newline
		\letter{C}
		\hspace{-0.5cm}
		\begin{minipage}[t]{0.90\colwidth}
		\flushleft
		\begin{subequations}
		\begin{equation}\label{eqn:fb_int_gain}
		g^2=\frac{\br{J_{vs}J_{wv}J_{xw}}^2}{H_xF(\omega)}
		\end{equation}
		\begin{equation}\label{eqn:fb_int_noise}
		N=\frac{\br{J_{xw}J_{wv}}^2\avgabssq{\eta_{v}}
+J_{xw}^2H_v\avgabssq{\eta_{w}}+F(\omega)\avgabssq{\eta_x}
}{H_xF(\omega)}
		\end{equation}
		\begin{equation}\label{eqn:fb_int_gnr}
		\frac{g^2}{N}=\frac{\br{J_{vs}J_{wv}J_{xw}}^2}
{\br{J_{xw}J_{wv}}^2\avgabssq{\eta_{v}}
+J_{xw}^2H_v\avgabssq{\eta_{w}}+F(\omega)\avgabssq{\eta_x}},
		\end{equation}
		\end{subequations}
		where \begin{subequations}
		\begin{equation}
		F(\omega)=\omega^4+\br{\mu_{v}^2+\mu_{w}^2+2J_{vw}J_{wv}}\omega^2+
		\br{J_{vw}J_{wv}-\mu_{w}\mu_{v}}^2
		\end{equation}
		\begin{equation}
		H_i=\omega^2+\mu_i^2
		\end{equation}
		\end{subequations}
		\end{minipage}
		\hspace{1.3cm}
		\letter{D}
		\hspace{0.2cm}
		\begin{minipage}[t]{0.90\colwidth}
			\includegraphics[angle=-90,scale=0.28]{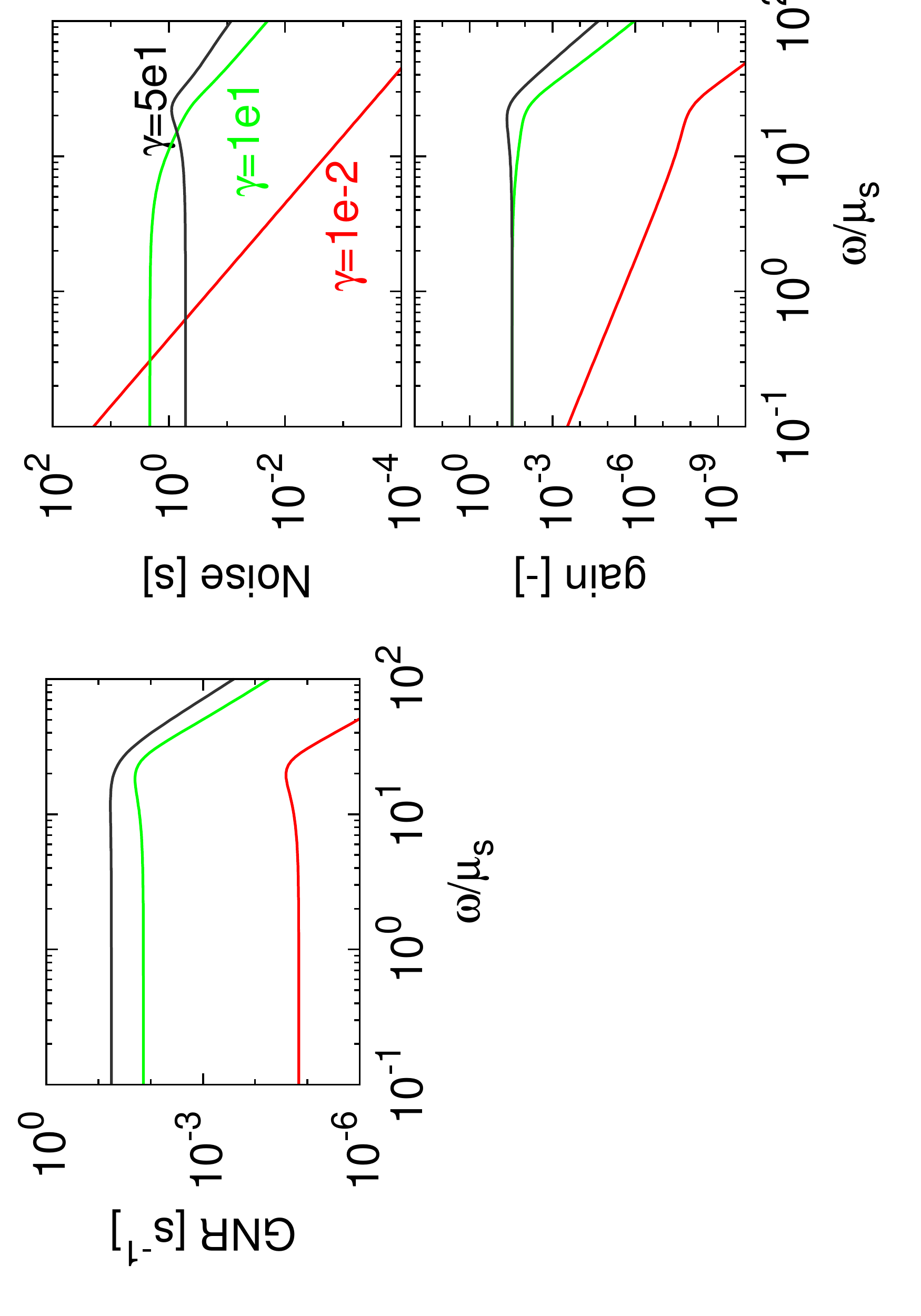}
		\end{minipage}
		\hspace{1.0cm}
			\letter{E}
		\begin{minipage}[t]{0.90\colwidth}
			\includegraphics[angle=-90,scale=0.28]{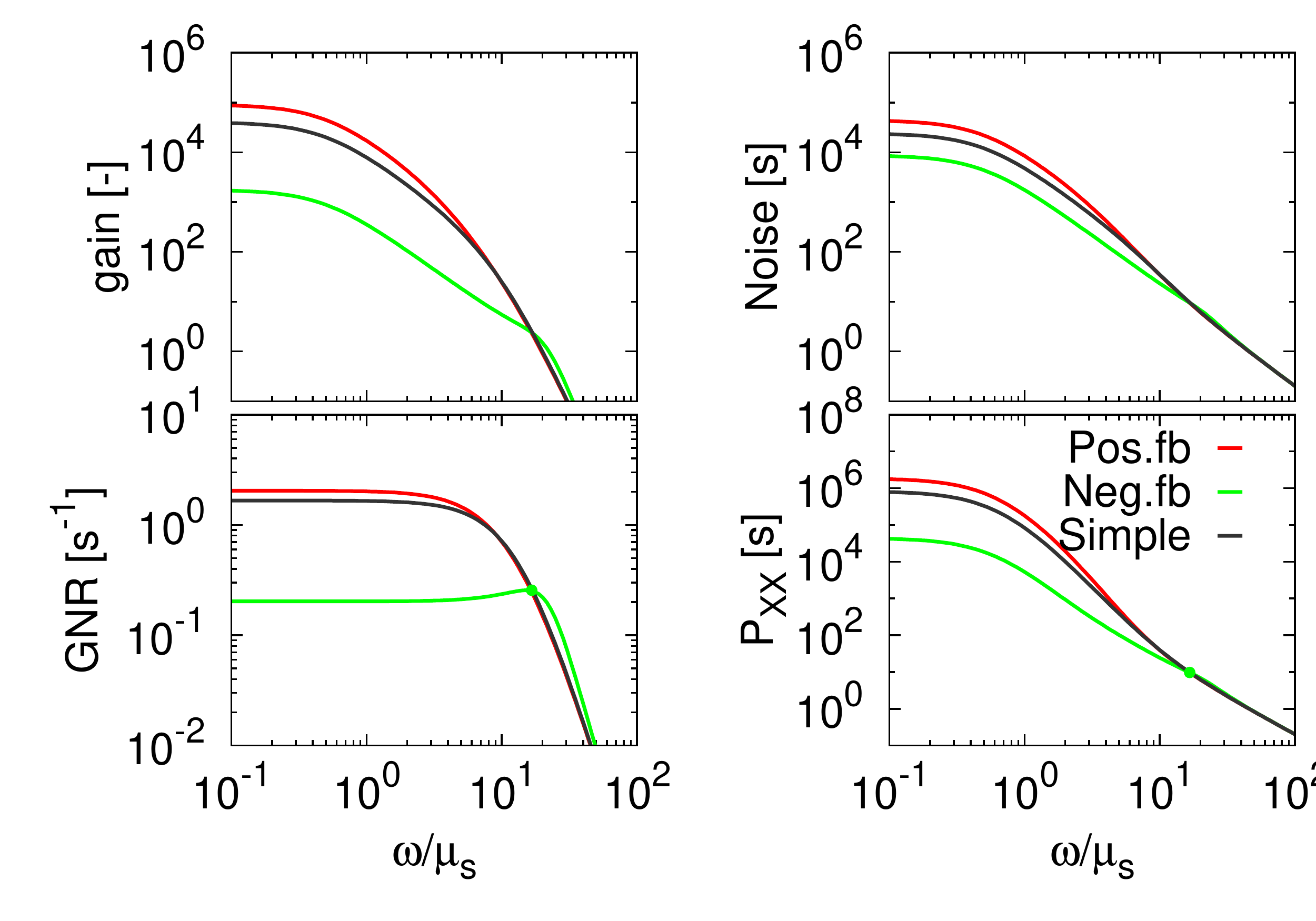}
		\end{minipage}
		\hspace{1.0cm}
		\letter{F}
		\hspace{0.2cm}
		\begin{minipage}[t]{0.90\colwidth}
			\includegraphics[angle=-90,scale=0.28]{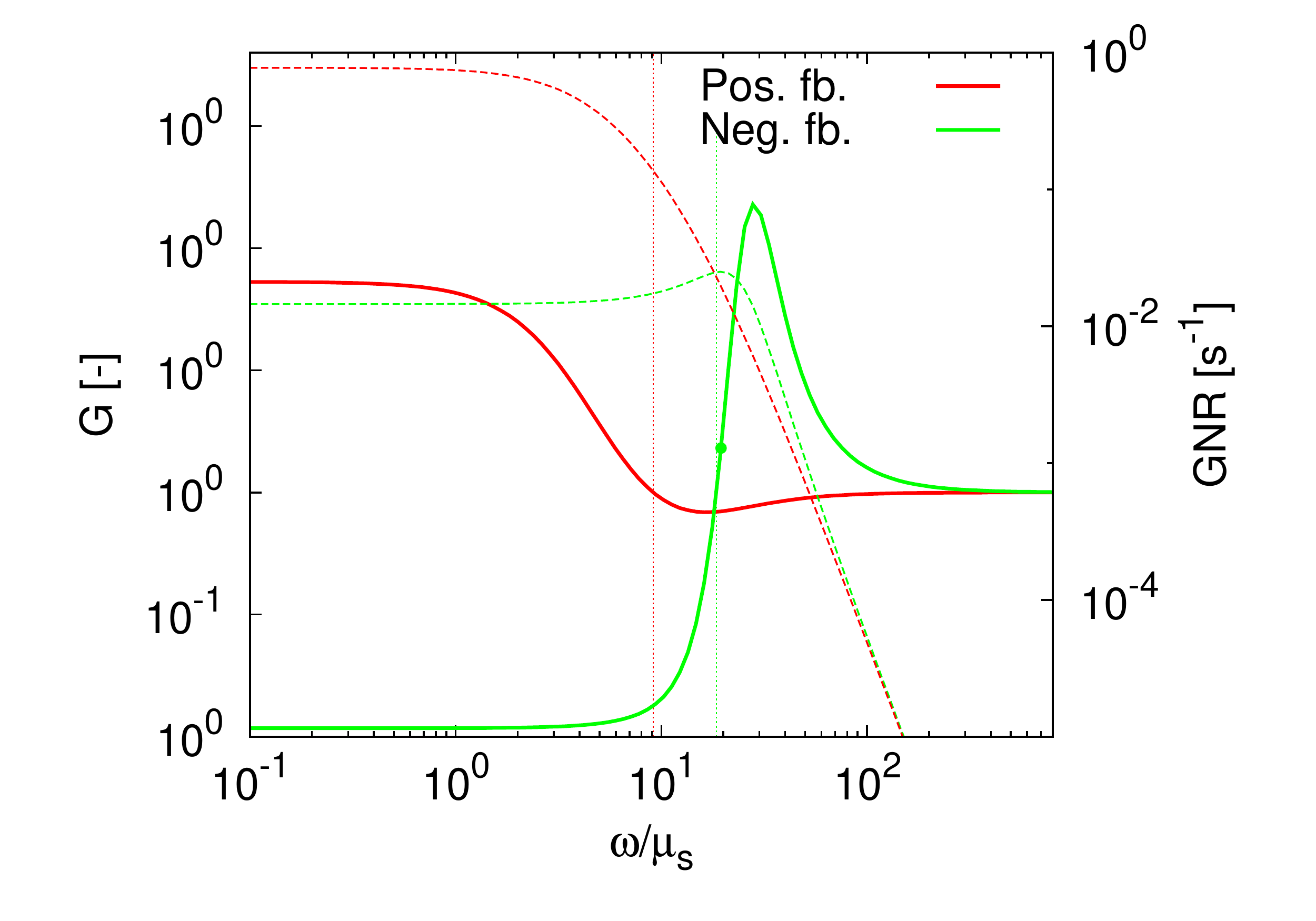}
		\end{minipage}
	\end{minipage}
}

	\caption{ {\bf A three-step cascade with feedback from an
			intermediate component, discussed in section {\em Feedback }}.
		\textbf{A}: Cartoon of a negative feedback motif, where $s$ is the signal
			and $x$ the response, and $w$ negatively regulates $v$.
		\textbf{B}: The Langevin equations of this motif.
		\textbf{C}: The characteristic equations: gain $g^2(\omega)$, noise
			$N(\omega)$ and gain-to-noise ratio $g^2(\omega) / N(\omega)$.
		\textbf{D}: The effect of changing the strength of the intrinsic noise in
			$x$, $N_x(\omega)$, on the spectra of the gain, noise, and gain-to-noise
			ratio of a cascade with negative feedback. $N_x(\omega)$ is varied by
			changing $\gamma(=J_{xw})$ and $\mu_x$, in such a way that
			$\langle x\rangle$ remains constant. Lines show: black, $\gamma=50$;
			green, $\gamma=10$; red, $\gamma=0.01$. Decreasing $\gamma$ and $\mu_x$
			in this way increases the relative contribution of $N_x(\omega)$ to
			the total noise. We see that as $\gamma$ is reduced the gain and noise
			decrease at frequencies $\omega<\mu_x$, but the noise increases at lower
			frequencies. The gain-to-noise ratio decreases at all frequencies.
			However, the peak in the gain-to-noise ratio becomes more pronounced.
			Parameters: $k_s=10$, $\beta=10$, $\nu=330$, $K=0.5\avg{w}$, $n=5$,
			$\mu_v=10$, $\mu_w=10$.
		\textbf{E}: The spectra of the gain, noise, gain-to-noise ratio, and the
			output power, $P_{xx}(\omega)$. For small $\omega$, positive feedback (red
			line) enhances the gain, noise, and gain-to-noise ratio, while negative
			feedback (green line) decreases these. For higher frequencies, negative
			feedback increases the gain, enhancing the gain-to-noise ratio. With
			negative feedback a peak in the gain-to-noise ratio is present (denoted by
			the green dot), while none is present in the output power spectrum
			$P_{xx}(\omega)$. Parameters: $k_s=10$, $\mu_w=10$, $\mu_v=10$,
			$\mu_x=0.5$, $\beta=10$, $\gamma=10$. For positive feedback:
			$K=0.5\avg{w}$, $n=1$ and $\nu=150$. For negative feedback:
			$K=0.5\avg{w}$, $n=4$ and $\nu=1700$.
		\textbf{F}: Solid lines show $G_{\rm fb}(\omega)$ (left axis), the
			gain-to-noise ratio for networks with positive (red) or negative (green)
			feedback divided by that of the corresponding simple cascade. Relative
			to the simple cascade, positive feedback increases the gain-to-noise
			ratio at low frequencies, while negative feedback increases the
			gain-to-noise ratio at high frequencies. Vertical lines indicate the
			frequencies at which $G_{\rm fb}(\omega)=1$ (Eqn.~\ref{eqn:fb_omega_switch}).
			Dashed lines show the gain-to-noise ratios for the positive and negative
			feedback motifs (right axis). Parameters: $k_s=10$, $\mu_w=10$,
			$\mu_v=10$, $\mu_x=1$, $\beta=10$, $\gamma=1$ and $K=0.5\avg{w}$. For
			positive feedback: $n=1$ and $\nu=150$. For negative feedback $n=5$
			and $\nu=3300$.\label{fig:fb_int}
	}

\end{figure*}

In section {\em Autoregulation} we saw that the
gain-to-noise ratio is sensitive to the precise position of autoregulation in
a cascade. In this section we therefore study a cascade where the feedback is
not from $x$ to $v$, but between two intermediate components $w$ and $v$ (see
Fig.~\ref{fig:fb_int}A and Eqn.~\ref{eqn:fb_int_net} in Fig.~\ref{fig:fb_int}B). This also
corresponds to taking the output of the previous feedback cascade (Fig.~\ref{fig:fb_res}A)
as the input to another downstream process.

Expressions for the gain, noise and gain-to-noise ratio are given in
Fig.~\ref{fig:fb_int}C. For positive feedback the gain, noise and gain-to-noise
ratio are once again monotonically decreasing with increasing frequency.
However, we find that for a network with strong negative feedback (Hill
coefficient $n>1$, see Eqn.~S1-54), the gain-to-noise ratio can have a maximum as a
function of frequency at
\begin{multline} \label{eqn:wopt}
	\omega_{\rm peak}^2=-\frac{1}{2}\br{
		\overbrace{J_{xw}^2\frac{\avgabssq{\eta_w}}{\avgabssq{\eta_x}}}
			^{\text{noise}}
		+\overbrace{\mu_v^2+\mu_w^2+2J_{vw}J_{wv}}
			^{\text{resonance}\br{\omega^2_{res}}}
		} \\
	=-\frac{1}{2}\br{
		J_{xw}^2\frac{\avgabssq{\eta_w}}{\avgabssq{\eta_x}}
		+\mu_v^2+\mu_w^2-2\frac{n\mu_w\mu_v\avg{w}^n}{K^n+\avg{w}^n}
	}.
\end{multline}
This peak frequency depends on the characteristic resonance frequency of the
feedback loop, $\omega_{\rm res}$, which is determined by the interactions between
$v$ and $w$: $\mu_v$, $\mu_w$, $J_{vw}$ and $J_{wv}$. It is additionally
dependent on the relative strengths of the noise introduced into the network at
$w$ and at $x$.

We can understand the appearance of this peak as follows. For a network with
negative feedback, $g^2(\omega)$ (Fig.~\ref{fig:fb_int}D, bottom right) has a maximum as a function
of frequency at $\omega_{\rm res}$, the characteristic resonance frequency of the feedback loop.
Input signals at this frequency are amplified by the constructive interference
between the signal transmitted to $v$ from $s$ and the signal which is fed back
from $w$ to $v$. We note that the resonance frequency has the same form as
Eqn.~\ref{eqn:omega_res}, and depends only on the interactions between $v$ and
$w$. The behaviour of the noise power spectrum (Fig.~\ref{fig:fb_int}D, top right) is more
complex. We consider two limiting cases in which different noise terms dominate. When the total
noise is dominated by noise introduced at $v$ or $w$, the noise is processed through
the feedback loop together with the signal. As discussed in the previous
section, $N(\omega)$ therefore shows a peak at a similar frequency to the gain (black line).
These two peaks cancel, and hence the gain-to-noise ratio (Fig.~\ref{fig:fb_int}D, top left, black
line) is monotonically decreasing with frequency. On the other hand, when the total noise is
dominated by $N_x(\omega)$ (top right, red line) the noise in the network is not affected by the
feedback loop. Hence no peak is found in the noise power spectrum. In this limit, the peak in
the gain-to-noise ratio corresponds to the peak in the gain at $\omega_{\rm res}$ (top left, red
line).

From these arguments we see that the peak in the gain-to-noise ratio becomes
more pronounced as the relative contribution of $N_x(\omega)$ to the total noise
increases. Additionally, increasing the strength of
the negative feedback by reducing $K$ or increasing $n$ leads to a more
pronounced peak. However, this increase in the {\em relative} peak height comes
at the expense of a reduction in the value of the gain-to-noise ratio at all
frequencies.

How does the gain-to-noise ratio of the network with feedback compare to the
corresponding (four-component) simple cascade? We examine the ratio of the
gain-to-noise for the network with feedback to the gain-to-noise of the simple
cascade,
\begin{equation}
	G_{\rm fb}(\omega)=\left[\frac{g^2(\omega)}{N(\omega)}\right]_{\rm fb}/
		\left[\frac{ g^2\br{ \omega}}{N(\omega)}\right]_{\rm simple},
\end{equation}
and find that (Fig.~\ref{fig:fb_int}E,F)
\begin{subequations} \label{eqn:fb_omega_switch} \begin{eqnarray}
	G_{\rm pos}(\omega) >1 &\text{ if }&
		\omega^2<\mu_v\mu_w\left[1-\frac{n}{2}\frac{K^n}{K^n+\avg{w}^n}\right]
		\label{eqn:fb_omega_switch_pos}, \\
	G_{\rm neg}(\omega) >1 &\text{ if}& \omega^2>
		\mu_v\mu_w\left[1+\frac{n}{2}\frac{\avg{w}^n}{K^n+\avg{w}^n}\right]
		\label{eqn:fb_omega_switch_neg}.
\end{eqnarray} \end{subequations}
Interestingly, for both types of feedback there is a range of frequencies over
which the gain-to-noise ratio increases relative to the simple cascade. This
contrasts to the results of section {\em Autoregulation}, where we found that
autoregulation affected the gain-to-noise ratio in the same way at all
frequencies.

This difference can again be understood in terms of the interference of the two
signals arriving at $v$. As described above (and in Eqn.S1-43-S1-44), at low frequencies
the signal propagated from $s$ to $v$ and the feedback signal from $w$ to $v$
are in phase, while at high frequencies the two signals are exactly out of
phase. Hence for a positive feedback loop (Fig.~\ref{fig:fb_int}E,F; red line) the signals combine
constructively at
low frequencies, increasing the gain, but destructively at high frequencies,
decreasing the gain. Recall that, since we are comparing networks with equal
production, the noise strengths $\avgabssq{\eta_v}$, $\avgabssq{\eta_w}$ and
$\avgabssq{\eta_x}$ are equal in the regulated and simple cascades. In an
analogous way to the autoregulation discussed in section {\em Autoregulation},
the presence of feedback between $w$ and $v$ affects both the signal and noise
introduced upstream of $x$, but not noise introduced at $x$. Hence, at low
frequencies positive feedback amplifies the signal and the noise introduced
at the levels of $v$ and $w$, but not noise introduced at $x$. Hence at low
frequencies the gain-to-noise ratio increases relative to the simple cascade. At
high frequencies, however, positive feedback reduces the gain and the noise
upstream of $x$, but not the intrinsic noise $N_x(\omega)$; consequently, the
gain-to-noise ratio is reduced compared to the simple cascade. Conversely, a
network with negative feedback (Fig.~\ref{fig:fb_int}E,F;green line) reduces the gain at low
frequencies, reducing the gain-to-noise ratio. However, at high frequencies, the feedback amplifies
the signal but not $N_x(\omega)$, leading to an increase in the gain-to-noise ratio.

From these results we conclude that if a cell is only concerned with low
frequency input signals, it is beneficial in terms of information transmission
to add positive feedback within the signalling cascade. If the system wishes to
respond specifically to high-frequency signals, negative feedback can be used to
increase the fidelity of transmission for these signals. Additionally for a
strong negative feedback ($n\gg1$ or $K\ll\langle w\rangle$, see Eqn.~S1-61) the
gain-to-noise can have a peak in the regime where signalling is more reliable
than for a simple cascade, allowing the cell to focus on signals in a particular
frequency band. We note that the negative feedback motifs considered here do not
lead to perfect adaptation to constant input signals, which is characterised by
$g^2(\omega=0)=0$ and is necessary for true band-pass behaviour. Perfect adaptation
requires that the feedback to be implemented via a buffer node or side branch
\cite{ma09}. An example of this network architecture is the {\em E. coli}
chemotaxis pathway \cite{yi00}, for which the gain-to-noise ratio does indeed
indicate a band-pass for information \cite{tostevin09}.

\section{Discussion}
In this paper we have analysed information transmission through a number of
network motifs, namely cascades, autoregulation and feedback. One of the most
important conclusions of our analysis is that to understand how reliably
biochemical networks can transmit time-varying signals, we have to study the
frequency-dependent gain-to-noise ratio \cite{tostevin09}. In particular, the
power spectrum of the output signal may not be a good measure for how
biochemical networks transduce time-varying input signals. The power spectrum of
the output signal depends on the power spectrum of the input signal, the
frequency-dependent gain, and the frequency-dependent noise. Only the latter two
quantities are intrinsic properties of the network, provided that the network
detects the input via biochemical reactions that do not affect the statistics of
the input signal \cite{tanasenicola06}. Moreover, we have seen that the power
spectrum of the output signal may differ qualitatively from that of the
frequency-dependent gain-to-noise ratio. A striking example is provided by the
network with negative feedback from the output component, which shows a peak in
the output signal (see Fig.~\ref{fig:fb_res}D): while one might be tempted to
conclude that input signals at this frequency are transduced most reliably, our
analysis shows that this peak in the output spectrum is simply the result of
resonant amplification of both the input signal and the noise in the network.

Our analysis leads us to draw the following conclusions on the effect of
autoregulation and feedback on the transmission of time-varying signals: 1)
autoregulation of the output component does not affect the gain-to-noise ratio,
and hence does not affect information transmission (Fig.~\ref{fig:autoreg_res}C);
2) positive autoregulation of an intermediate component increases the
gain-to-noise ratio over all frequencies, while negative autoregulation tends to
decrease it over all frequencies (Fig.~\ref{fig:autoreg_int}D); 3) negative
feedback from the output component onto an upstream component may lead to a peak
in the power spectrum of the output, and those of the gain and the noise; yet,
even though the peaks of gain and the noise can be at different frequencies,
negative feedback from the output component onto an upstream component can {\em
not} lead to a peak in the spectrum of the gain-to-noise ratio
(Fig.~\ref{fig:fb_res}D); 4) positive feedback between upstream components
enhances the gain-to-noise ratio at low frequencies, while negative feedback
increases the gain-to-noise ratio at high frequencies (Fig.~\ref{fig:fb_int}F).
Further, we note that it is possible to achieve a peak in the gain-to-noise
ratio via negative feedback between components that are upstream of the output
component (Fig.~\ref{fig:fb_int}D); however, this comes at the expense of a
reduction in the gain-to-noise ratio for all frequencies. We also note here that
stronger band-pass filtering of information can be obtained with networks employing integral
feedback in a side branch\cite{tostevin09}, as found in the networks of osmo adaptation
\cite{mettetal08} or bacterial chemotaxis \cite{yi00}. Alternatively, band-pass
filters for information transmission can be obtained via feed{\em forward}
loops, which we will discuss in a forthcoming publication.

\begin{figure*}[!ht]
	\includegraphics[angle=-90,scale=0.7]{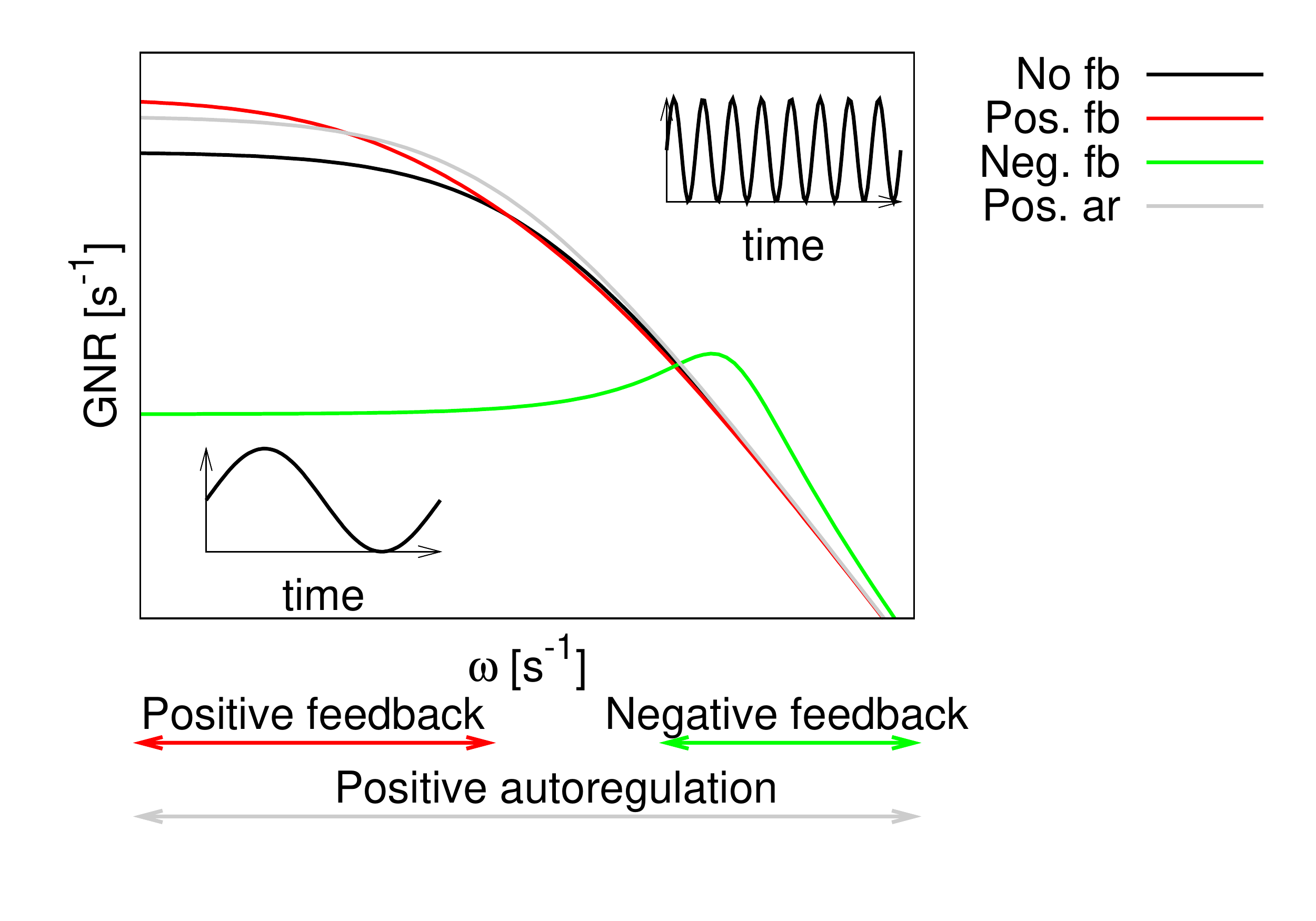}
	\caption{{\bf A schematic drawing of the main conclusions of this paper}. The
frequency of the variations of the input is shown on the x-axis. For three different motifs the
gain-to-noise ratio is shown.  The arrows indicate the specific frequency regime for which each
motif performs better with respect to a simple cascade.\label{fig:cartoon}}
\end{figure*}

Taken together these results reveal the following design principles for the use
of feedback and autoregulation in signal transduction cascades (see the schematic
drawing Fig. \ref{fig:cartoon}). Firstly, feedback and autoregulation can improve
information transmission, but only if they occur upstream of the dominant source
of noise in the cascade. Feedback or autoregulation downstream of the dominant
noise source affects the gain and the noise similarly. Secondly, if signals over
the full frequency range have to be transmitted reliably, positive
autoregulation is advantageous, while if the cell is concerned only with low-
or high-frequency signals, then positive or negative feedback can be employed.

The approach employed here has a number of limitations. Firstly, we have used
the linear-noise approximation, and the power spectra calculated using this
approximation may deviate from those of the full non-linear system. We argue
that this effect does not significantly affect our results, since we find
excellent agreement between the power spectra calculated analytically using the
linear-noise approximation and those obtained from stochastic simulations of the
full system (see S1).
The second potential source of inaccuracy is the use of
Eqns.~\ref{eqn:rate_transmission} and \ref{eqn:rate_SN}, which are exactly only
for linear Gaussian systems. However, the information rate calculated in this
approximation provides a lower bound on the information transmission rate of the
full system \cite{mitra01}.

Another limitation of our analysis is that to reduce the complexity of the
problem, we have assumed that the networks obey the spectral-addition rule
\cite{tanasenicola06}, meaning that reactants are not consumed during reaction
events. However, irreversible modifications of a substrate molecule are common
in biochemical networks, and reactions of this type can significantly change the
correlations between different network components. For instance, in a cascade of
the type $X_0\to X_1\to\dots X_{n-1}\to X_n$, where in each reaction step the
reactant is consumed, correlations of the form $\avg{\eta_i\eta_{i+1}}
=-k\avg{X_i}$ appear between different noise terms. As a result, for this
cascade the covariance between different components $\avg{x_ix_{j\neq i}}=0$
\cite{tanasenicola06, levine07}, and hence the mutual information between {\em
instantaneous} levels of components $X_i$ and $X_{j\neq i}$ is zero
\cite{tostevin09}. This may suggest that these cascades cannot effectively
transmit information. Yet, the analysis of \cite{tostevin09} indicates that this
motif can, in fact, reliably transmit time-varying signals. It would therefore
be of interest to study the effect of cross-correlations in the noise on the
information transmission in the motifs studied here. We leave this for future
work.

Lastly, how could our predictions be tested experimentally? It is increasingly
being recognised that stimulating biochemical networks with time-varying signals
provides a wealth of information on the dynamics of these networks
\cite{lipan05, austin06, tan07, mettetal08,tu08, cournac09}. These experiments
can also be used to study the reliability by which biochemical networks can
transmit time-varying signals. By measuring not only the power spectra of the
in- and output signals, $P_{ss}(\omega)$ and $P_{xx}(\omega)$, but also their
cross-power spectrum $P_{sx}(\omega)$, one can obtain the frequency-dependent
gain $g^2(\omega) \equiv \abs{P_{sx}(\omega)}^2/P_{ss}(\omega)^2$ and the
frequency-dependent noise $N(\omega)$ (see Eqn. \ref{eqn:Sadd}), and hence the
gain-to-noise ratio. Stimulating synthetic gene circuits or existing signal
transduction pathways and gene regulation networks of known architecture with
time-varying signals, for example using microfluidic devices, would make it
possible to test our predictions on the effect of feedback and autoregulation on
information transmission.

\section{Acknowledgments}
We thank Martin Frimmer and Nils Becker for a careful reading of this manuscript. This work is part
of the research program of the ``Stichting voor Fundamenteel Onderzoek der Materie (FOM)'', which is
financially supported by the ``Nederlandse organisatie voor Wetenschappelijk Onderzoek (NWO)''.

\bibliography{arxiv_paper_fb_v1}

\setcounter{figure}{0}
\setcounter{equation}{0}
\setcounter{section}{0}
\renewcommand{\theequation}{S1-\arabic{equation}}
\renewcommand{\thefigure}{S1-\arabic{figure}}\renewcommand{\figurename}{Figure}
\renewcommand{\thesection}{S1-\arabic{section}}
\clearpage
\newpage

\onecolumngrid
\section{S1}
All cascades have the following simple (linear) birth-death process  for the signal
\begin{equation}
\frac{ds}{dt}=k_s-\mu_s s+\Gamma\br{t}
\end{equation}

\section{Gillespie Simulations}
The linearization used in the derivation can change the characteristics of the frequency
response. A linearized system does not change the frequency of the transmitted signal. However, this
may not be the case for a non-linear system.  To study this, we
performed Gillespie simulations of the full system. The positive and negative regulation in our
networks arises from Hill-like interactions between components. In the Gillespie simulation we
calculated the propensities for every reaction with identical expressions. For example, in the
network with negative feedback from $w$ to $v$, we model reactions like Eqn.~1a in
\ref{fig:fb_int}B as
\begin{equation}
S \xrightarrow{r} S + V
\end{equation}
where $r$ is
\begin{equation}
r=\frac{\nu K^n s}{K^n+w^n}.
\end{equation}
In these equations the actual copy number $w$ is used, and not $\avg{w}$, as in the linearized
expressions (Eqns.~2a,2b in Fig.~\ref{fig:fb_int}C).\\

The power spectra are calculated using $2^{11}$ (2048) exponentially distributed frequencies from
$\omega=10^{-3}$ to $\omega=10^3$ and averaged over $2^4$ neighbouring frequencies to obtain a
single data point. In total we have $2^7$ datapoints. The length of the simulation is $10^6$
seconds, or a maximum of $10^9$ events. For every run 50 blocks are averaged.\\

\begin{figure*}[!ht]
\begin{tabular}{lrlr}
\letter{A} & \includegraphics[angle=-90, scale=0.2]{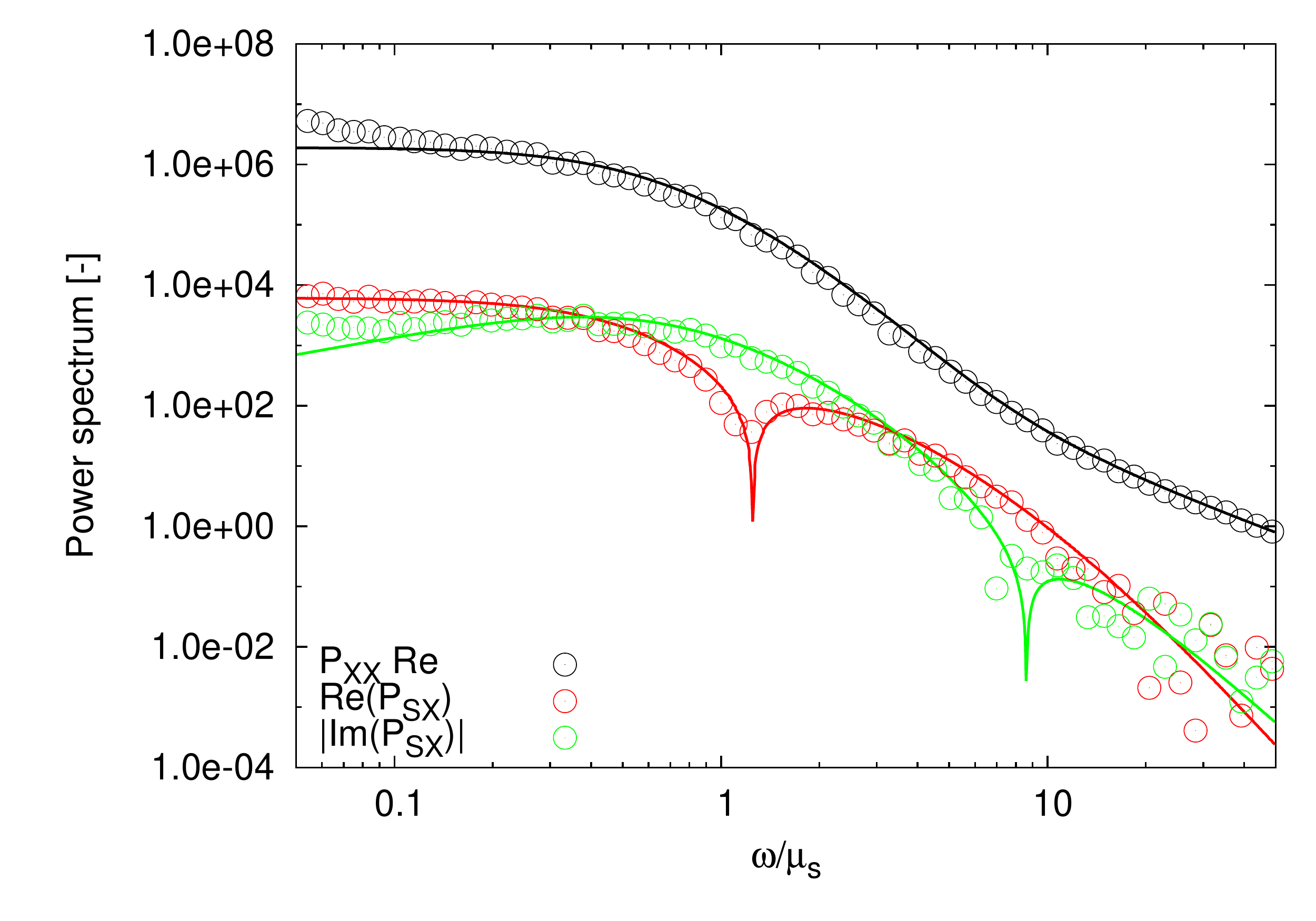} &
\letter{B} & \includegraphics[angle=-90, scale=0.2]{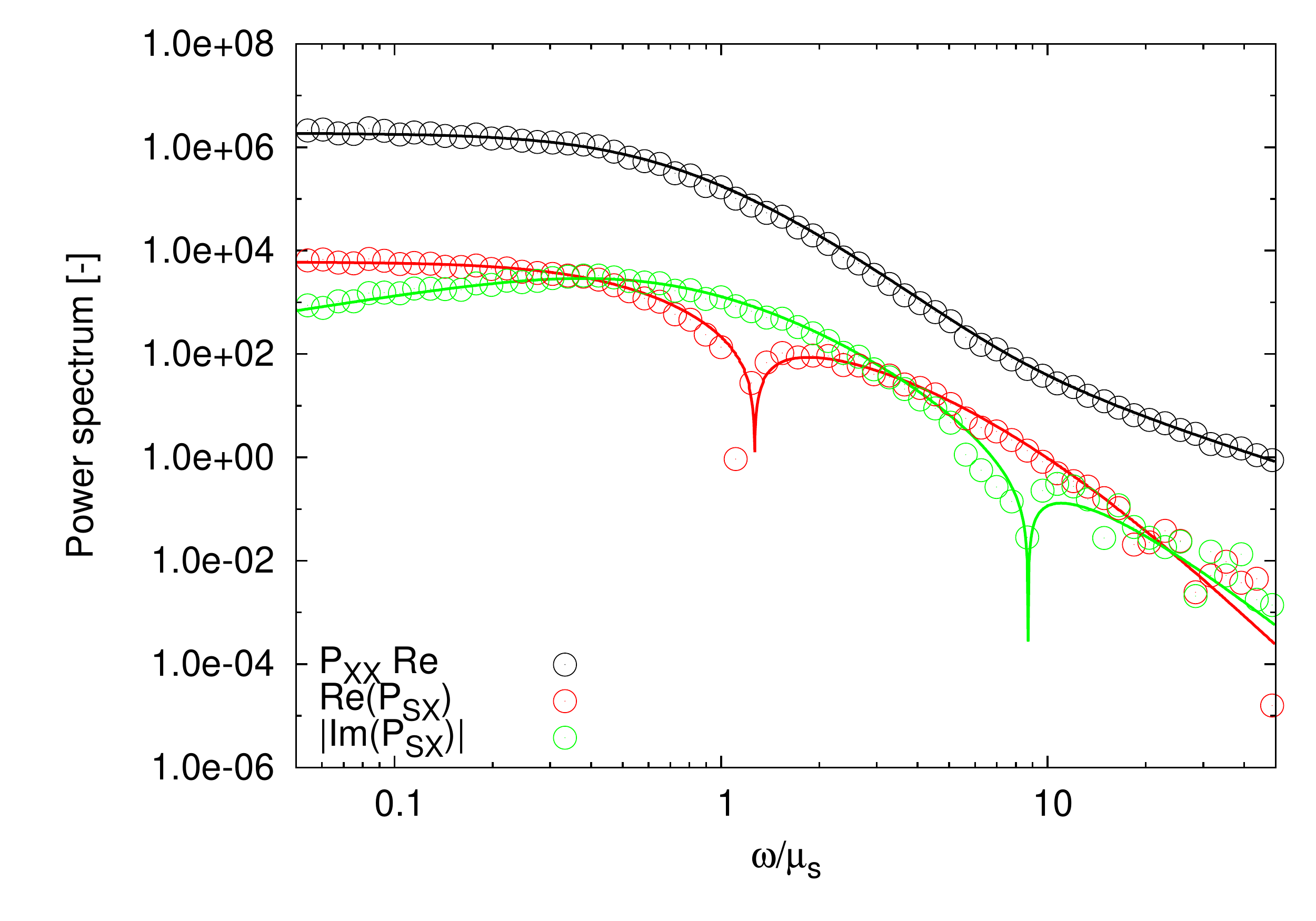} \\
\letter{C} & \includegraphics[angle=-90, scale=0.2]{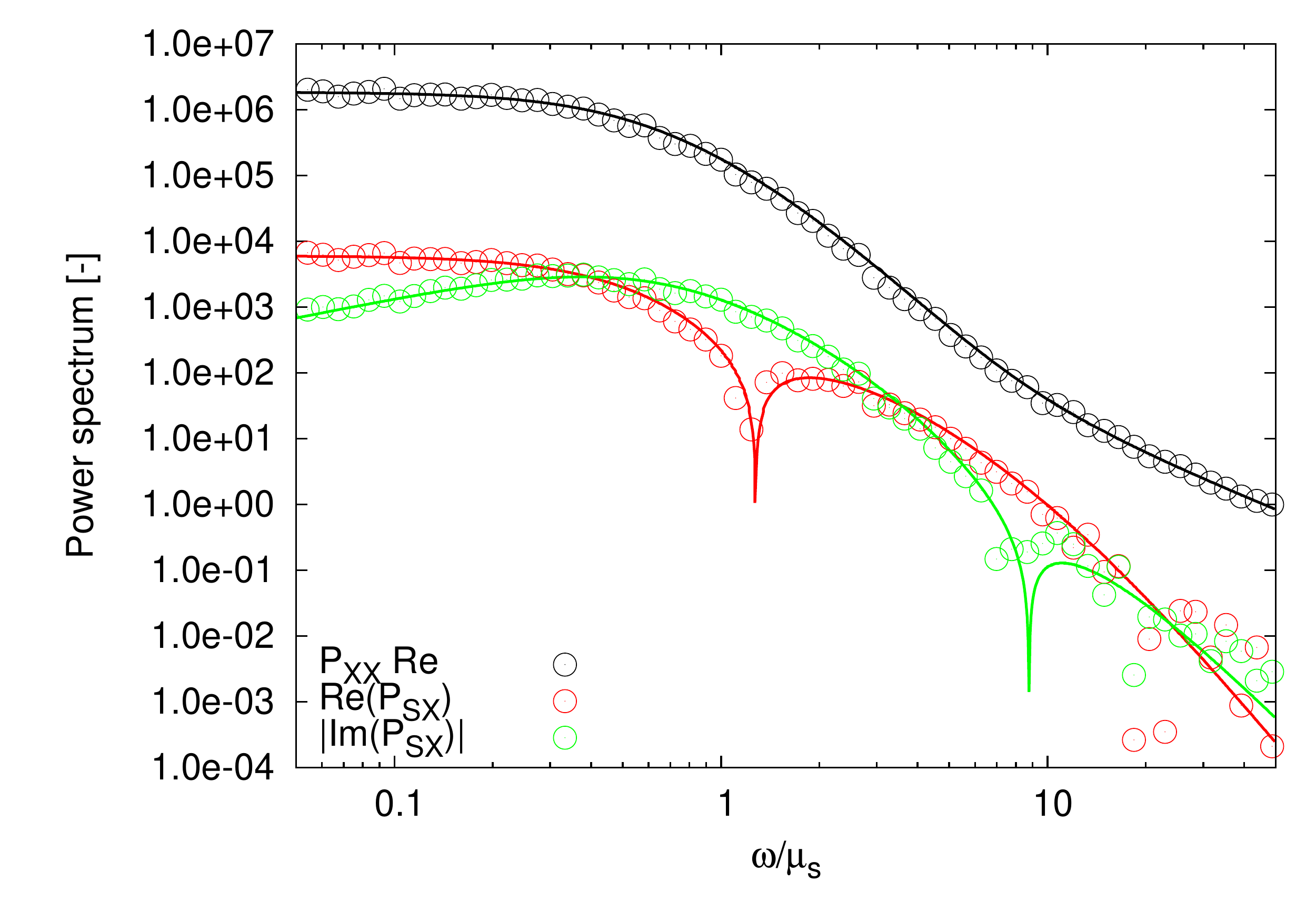} & &
\end{tabular}
\caption{\label{fig:si:a01} The results (circles) of the Gillespie simulations for a network with
positive feedback from $w$ to $v$ (Eqn.\ref{eqn:fb_int_net} in Fig.~\ref{fig:fb_int}B), together
with the results of
the
linear noise approximation (solid lines) as employed in the main text. Kinetic rates as
in Fig.~\ref{fig:fb_int}D with positive feedback. To drive the system to the non-zero steady state,
basal production of $w$ is present (Eqn.~\ref{eqn:si:fb_int_pos}),
\textbf{A})  with $a=0.1$,
\textbf{B})  with $a=1$,
\textbf{C})  with $a=10$}
\end{figure*}

The positive feedback loops considered here display bistability. For the positive feedback loops
a constant low level production is added to drive the system to the stable state
with high copy numbers, instead of the stable state where the copy number equals zero. For the
positively autoregulated component this is described by that
\begin{equation} \label{eqn:si:autoreg_pos}
\frac{dv}{dt}=-\mu_v v+\eta_v + \begin{cases} \frac{\nu vs}{K+v} \text{ if } v\neq 0 \\
\frac{1}{1000} \text{ if } v=0 \end{cases}
\end{equation}
Linearizing this we find that the fluctuations follow
\begin{equation}
\frac{d\tilde{v}}{dt}=-\mu_v \tilde{v}+\eta_v - \frac{\nu K\avg{s}}{K+\avg{v}}\tilde{v} + \frac{\nu
K}{K+\avg{v}}\tilde{s} \end{equation}
which is equivalent to the linearization of Eqn.~1 in Fig.~\ref{fig:autoreg_int}A. The addition of
the basal expression therefore drives the system to a specific steady state, but does not change the
dynamic behaviour around this steady state.\\
For positive feedback within the cascade, the motif is described by
\begin{equation} \label{eqn:si:fb_int_pos}
\frac{dw}{dt}=a +k_w v -\mu_w w+\eta_w.
\end{equation}
Taking different values for $a=0.1, 1, 10$ does not lead to qualitatively different answers (see
Fig.~\ref{fig:si:a01}). Again, the basal production
changes the steady state, but not the dynamical behaviour around the steady state.

\begin{figure} [!ht]
\includegraphics[angle=-90, scale=0.2]{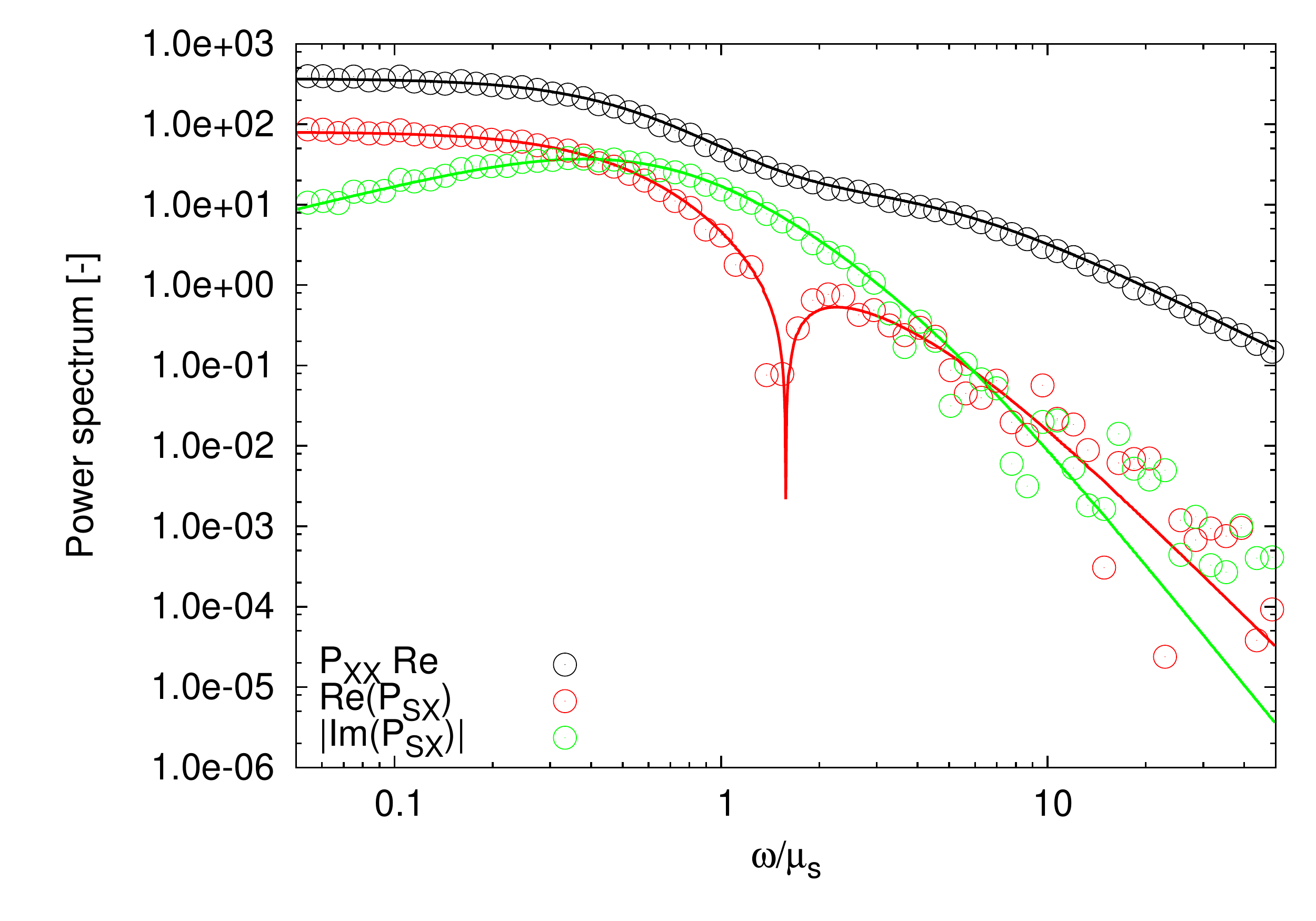}
\caption{\label{fig:si:148} The results (circles) of the Gillespie simulations for the linear
cascade (Eqn.~\ref{eqn:lin_casc}), together with the results of the linear noise approximation
(solid lines) as employed in the main text. Kinetic rates as in figure
\ref{fig:pow_spec}.}
\end{figure}

\begin{figure*}[!ht]
\begin{tabular}{lrlr}
\letter{A} & \includegraphics[angle=-90, scale=0.2]{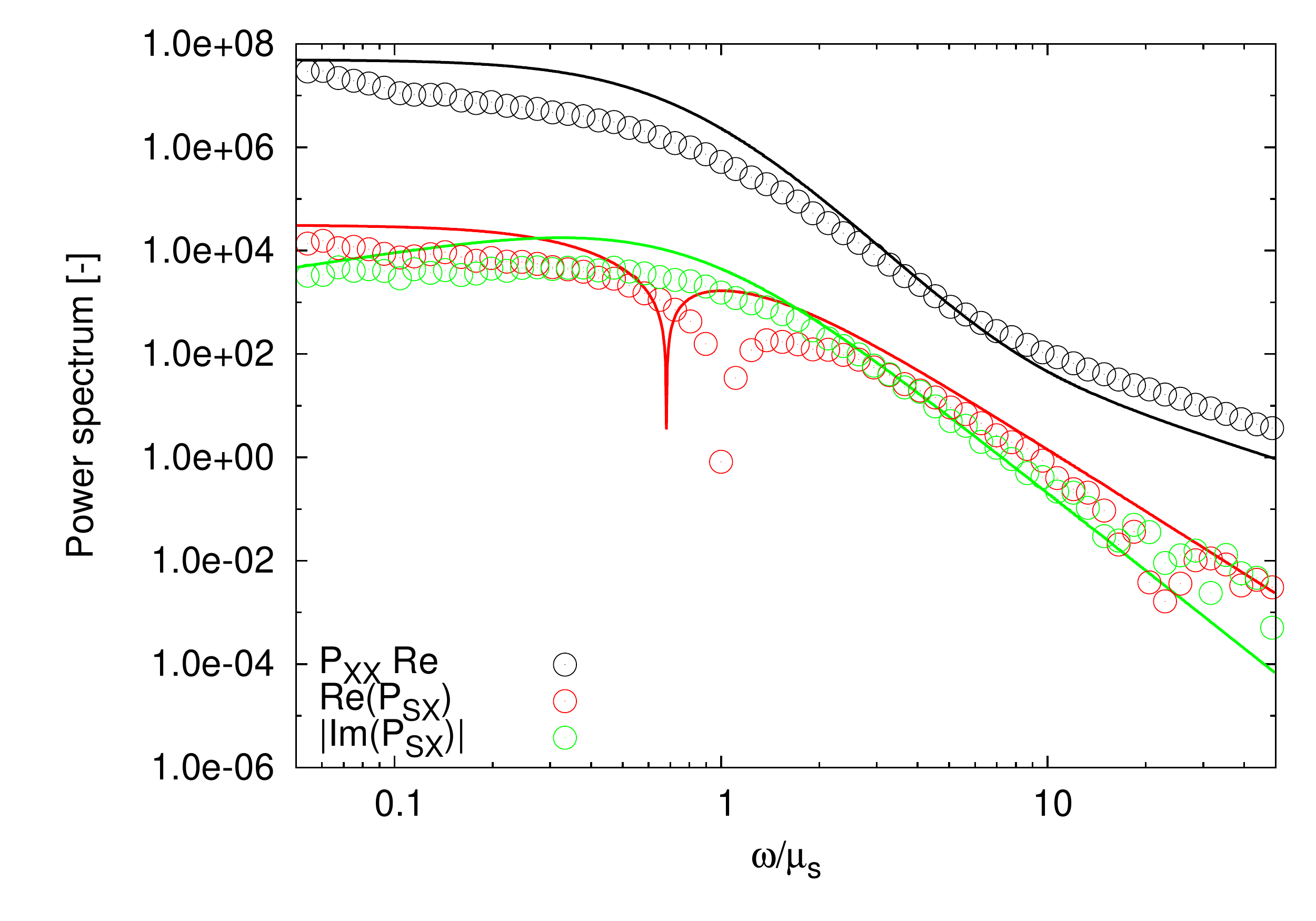} &
\letter{B} & \includegraphics[angle=-90, scale=0.2]{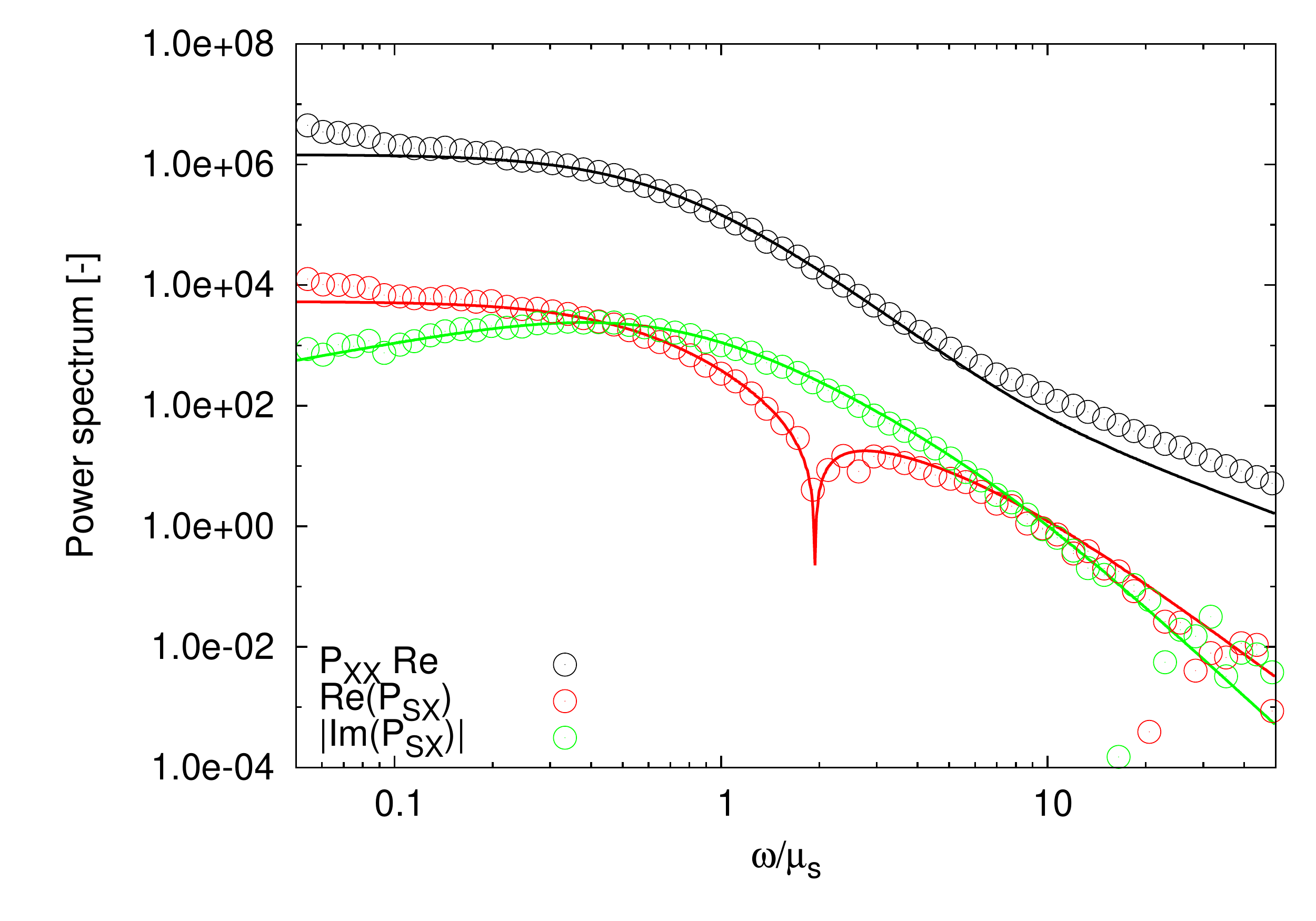}
\end{tabular}
\caption{\label{fig:si:150}
\textbf{A}) The results (circles) of the Gillespie simulations for the network with
positive autoregulation of $v$ (Eqn.\ref{eqn:autoreg_int_net} in Fig.~\ref{fig:autoreg_int}B)
(kinetic rates as in
Fig.~\ref{fig:autoreg_int}C with positive autoregulation), together with the results of the linear
noise approximation (solid lines) as employed in the main text.
To drive the system to the non-zero steady state, basal production of $v$ is present
(Eqn.~\ref{eqn:si:autoreg_pos}). The steady state of the full non-linear Gillespie simulation is
slightly different from the steady state derived from the mathematical expressions for $s, v$ and
$x$. This causes the slight difference between the results of the linearization and the
simulations.
\textbf{B}) The results (circles) of the Gillespie simulations for a network with
negative autoregulation on $v$ (Eqn.\ref{eqn:autoreg_int_net} in Fig.~\ref{fig:autoreg_int}B).
Together with the results
of the linear noise approximation (solid lines) as employed in the main text. Kinetic rates as
in Fig.~\ref{fig:autoreg_int}D with negative autoregulation.}
\end{figure*}

\begin{figure} [!ht]
\includegraphics[angle=-90, scale=0.2]{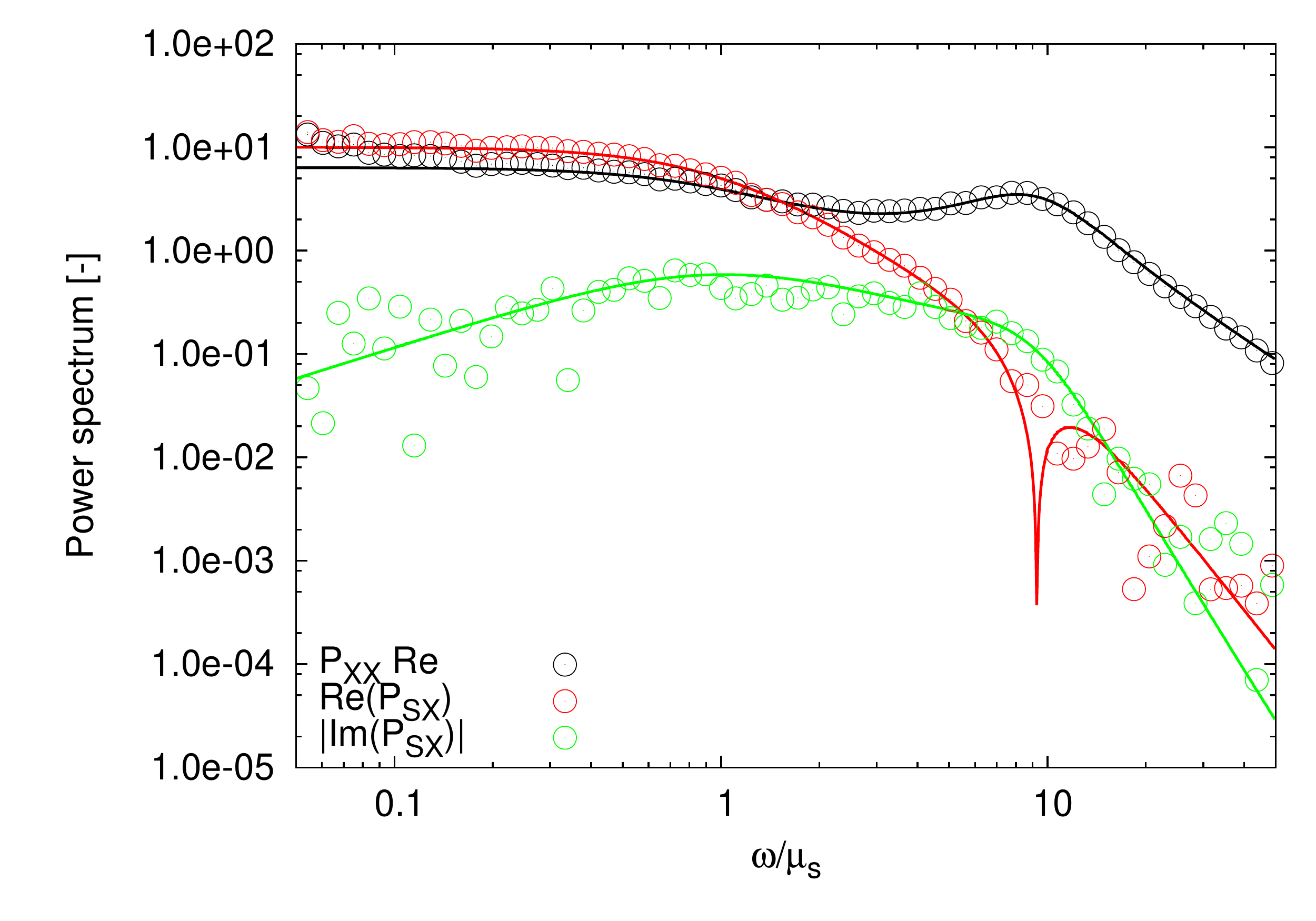}
\caption{\label{fig:si:151} The results (circles) of the Gillespie simulations for a network with
negative feedback from $x$ to $v$ (Eqn.\ref{eqn:fb_res_net} in Fig.~\ref{fig:fb_res}B),
together with the results of
the linear noise approximation (solid lines) as employed in the main text. Kinetic rates as
in Fig.~\ref{fig:fb_res}D.}
\end{figure}

\begin{figure} [!ht]
\includegraphics[angle=-90, scale=0.2]{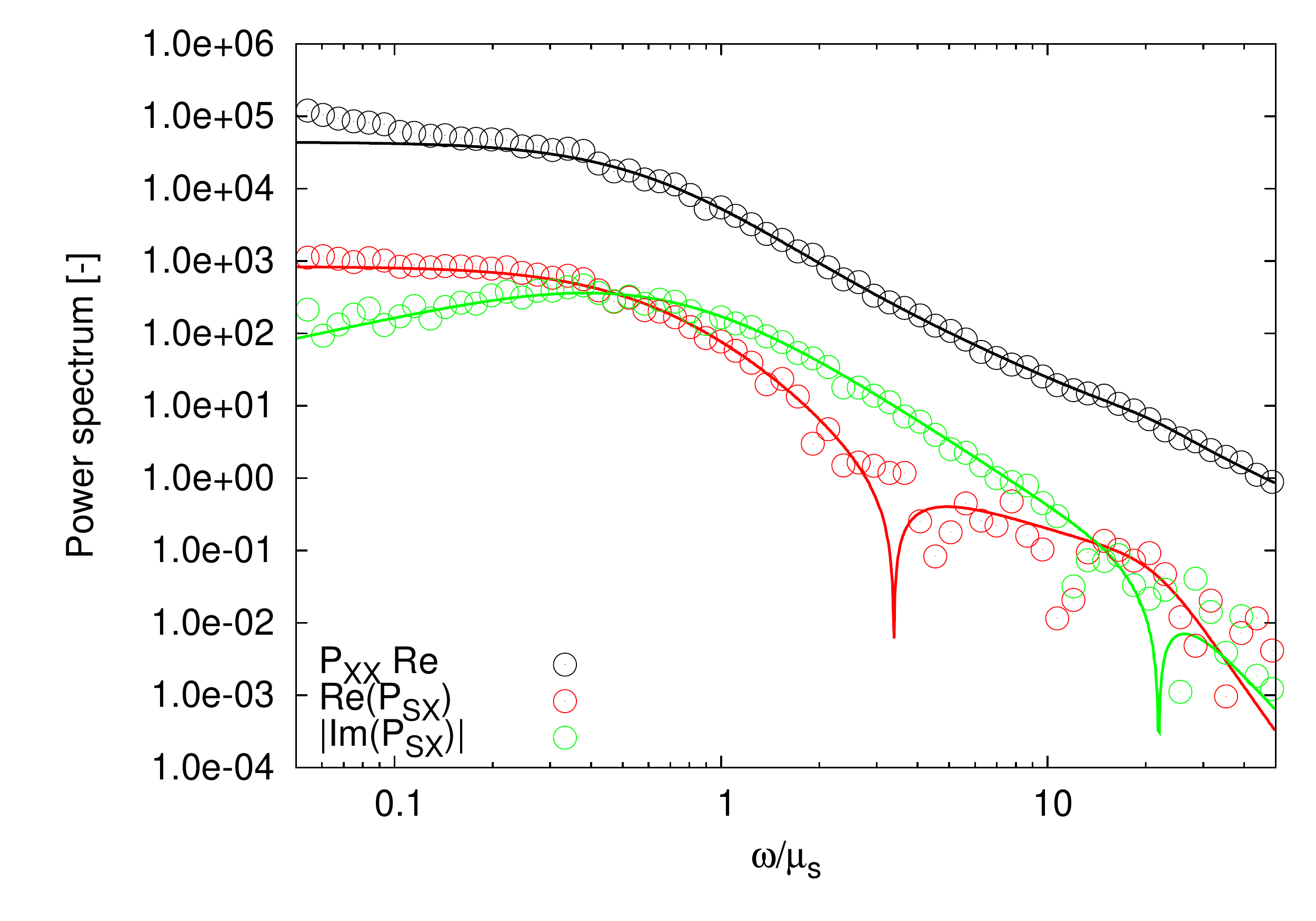}
\caption{\label{fig:si:153} The  results (circles) of the Gillespie simulations for a network with
negative feedback from $w$ to $v$ (Eqn.\ref{eqn:fb_int_net} in Fig.~\ref{fig:fb_int}B), together
with the results of
the linear noise approximation (solid lines) as employed in the main text. Kinetic rates as
in Fig.~\ref{fig:fb_int}D with negative feedback.}
\end{figure}
\clearpage

\section{Simple cascade}
The one step simple cascade is described by
\begin{equation}
\frac{dx}{dt}=k_x s - m_x x+\eta_x\br{t}
\end{equation}
with the following characteristic equations
\begin{subequations}
\begin{equation}
g^2\br{\omega}=\frac{k_x^2}{\omega^2+m_x^2}
\end{equation}
\begin{equation}
N\br{\omega}=\frac{\avgabssq{\eta_x}}{\omega^2+m_x^2}
\end{equation}
\begin{equation}\label{eqn:si:simple2s}
\frac{g^2}{N}=\frac{k^2_x}{\avgabssq{\eta_x}}
\end{equation}
Here, $\avgabssq{\eta_x}=k_x\avg{s}+m_x \avg{x}=2k_x\avg{s}$.
\end{subequations}
The three component simple cascade is described by (compare Eqn.~\ref{eqn:lin_casc})
\begin{subequations}
\begin{equation}
\frac{dv}{dt}=k_v s-m_v v+\eta_v\br{t}
\end{equation}
\begin{equation}
\frac{dx}{dt}=k_x v - m_x x+\eta_x\br{t}
\end{equation}
\end{subequations}
with the following characteristic equations
\begin{subequations}
\begin{equation}
g^2\br{\omega}=\frac{k_v^2k_x^2}{\br{\omega^2+m_x^2}\br{\omega^2+m_v^2}}
\end{equation}
\begin{equation}
N\br{\omega}=\frac{k_x^2\avgabssq{\eta_v}+\br{\omega^2+m_v^2}\avgabssq{\eta_x}}{\br{
\omega^2+m_x^2}\br{\omega^2+m_v^2}}
\end{equation}
\begin{equation}
\frac{g^2}{N}=\frac{k^2_vk^2_x}{k_x^2\avgabssq{\eta_v}+\br{\omega^2+m_v^2}\avgabssq{\eta_x}}
\end{equation}
\end{subequations}
The simple cascade is used as a reference. For the kinetic rates of the simple cascade we use roman
symbol ($k$ and $m$). For the kinetic rates of the cascades with feedback regulation we use greek
symbols.
\section{Autoregulation}
\subsection{Autoregulation by $x$}
An elementary network for autoregulation by $x$ onto itself is
\begin{subequations}
\begin{equation}
\frac{dx}{dt}=f\br{x}s - \mu_x x+\eta_x\br{t}, \text{where}
\end{equation}
\begin{equation}
f\br{x}s= \frac{\nu\kappa}{K+x}s
\begin{cases}
  \kappa=K, & \text{negative regulation} \\
  \kappa=x, & \text{positive regulation}
\end{cases}
\end{equation}
\end{subequations}
The gain, noise and gain-to-noise for this network are
\begin{subequations}
\begin{equation}
g^2\br{\omega}=\frac{J_{xs}^2}{\omega^2+J_{xx}^2}=
\frac{1}{\omega^2+\br{\mu_x-\pdiff{f\br{\avg{x}}}
{\avg{x}}\avg{s}}^2}f\br{\avg{x}}^2
\end{equation}
\begin{equation}
N\br{\omega}=\frac{\avgabssq{\eta_x}}{\omega^2+J_{xx}^2}
\end{equation}
\begin{equation}
\frac{g^2}{N}=\frac{J_{xs}^2}{\avgabssq{\eta_x}}=\frac{\br{\pdiff{f\br{
\avg{s},\avg{x}}}{\avg{s}}}^2}{\avgabssq{\eta_x}}
\end{equation}
\end{subequations}
For equal average production, as the simple three component cascade, (production rate $k_x$), we
chose
\begin{equation}
\avg{f\br{x}s}\equiv f\br{\avg{x}}\avg{s}=k_x\avg{s}
\end{equation}
where the first equation expresses the fact that we assume that the average rates can be
expressed by the rates at the deterministic steady state, thus ignoring fluctuations. Thus
\begin{equation}
k_x=J_{xs}=\pdiff{f\br{\avg{s},\avg{x}}}{\avg{s}}=\frac{\nu\kappa}{\br{K+\avg{x}}}
\end{equation}
and $\avg{\abs{\eta_x}^2}=2k_x\avg{s}$.
Expressed in terms of the kinetic rates of the simple cascade, the autoregulated cascade has the
following form
\begin{equation}
GNR=\frac{k_x^2}{2k_x\avg{s}}=\frac{k_x}{2\avg{s}}
\end{equation}
which is identical to Eqn.~\ref{eqn:si:simple2s}.
The power spectrum of $x$ for the autoregulated cascade is
\begin{equation}
P_{XX}\br{\omega}=\frac{J_{xs}^2\avgabssq{\Gamma}+\br{\omega^2+\mu_s^2}\avgabssq
{\eta_x}}
{\br{\omega^2+\mu_s^2}\br{\omega^2+\br{\mu_x-\pdiff{f\br{\avg{s},\avg{x}}}
{\avg{x}}}^2}}
\end{equation}
Following a rescaling of the kinetic degradation rate $\mu_x$, such that
$\mu_x^{\rm new}=\mu_x-J_{xx}$,  we observe that the power spectrum of the simple
cascade and the autoregulated cascade agree. This is because the noise term
$\eta_x$ depends on the mean rate of the production and degradation events. In steady state
the average number of production events equals the average number of degradation events. Since by
the rescaling the production is not changed, the noise $\eta_x$ is constant. The change in
$\mu_x\to\mu^{\rm new}_x$ will lead to a new steady state value $\avg{x}$, but not to a different
number of degradation events.
\subsection{Autoregulation by $v$}
For autoregulation of one of the intermediate components the network is
\begin{subequations}
\begin{equation}
\frac{dv}{dt}=f\br{v}s-\mu_v v+\eta_v\br{t}
\end{equation}
\begin{equation}
\frac{dx}{dt}=\beta v - \mu_x x+\eta_x\br{t}, \text{where}
\end{equation}
\begin{equation}
f\br{v}s= \frac{\nu\kappa}{K+v}s
\begin{cases}
  \kappa=K, & \text{negative regulation} \\
  \kappa=v, & \text{positive regulation}
\end{cases}
\end{equation}
\end{subequations}
The gain and noise for this network are
\begin{subequations}
\begin{equation}
g^2\br{\omega}=\frac{\br{J_{xv}J_{vs}}^2}{\br{\omega^2+\mu_x^2}\br{\omega^2+
J_{vv}^2}}
\end{equation}
\begin{equation}
N\br{\omega}=\frac{\beta^2\avgabssq{\eta_v}+\br{\omega^2+J_{vv}^
2}\avgabssq{\eta_x}}{\br{\omega^2+\mu_x^2}\br{\omega^2+
J_{vv}^2}}
\end{equation}
\end{subequations}
where $J_{vs}=f\br{\avg{v}}$. We equalize
the  production for $v$ between the autoregulated cascade and the three component simple cascade
(with rates $k_v \andd k_x$) to obtain
\begin{equation}\label{eqn:si:constr1}
k_v=J_{vs}=\frac{\nu\kappa}{\br{K+\avg{v}}}
\end{equation}
and the gain-to-noise ratio for the autoregulated cascade expressed in terms of the kinetic rates
of the simple cascade (thus using $k_v, k_x$ and $m_v,m_x$ where applicable) is
\begin{equation} \label{eqn:si:gnr_ar_vs_simple}
\frac{g^2}{N}=\frac{\br{J_{vs}J_{xv}}^2}{\beta^2\avgabssq{\eta_v}+\br{\omega^2+J_{vv}^
2}\avgabssq{\eta_x}}=\frac{k_v^2\beta^2}{\beta^2\avgabssq{\eta_v}+\br{
\omega^2+\br{\mu_v^2-\pdiff{f\br{\avg{s},\avg{v}}}{\avg{v}}}^2}\avgabssq{\eta_x}}.
\end{equation}
We keep all kinetic rates equal in the autoregulated and simple cascade that do not
influence the constraint condition (Eqn.~\ref{eqn:si:constr1}) $\br{\text{thus }\mu_v=m_v \andd
\beta=k_x}$. We then obtain
\begin{equation}
\frac{g^2}{N}=\frac{k_v^2k_x^2}{k_x^2\avgabssq{\eta_v}+\br{
\omega^2+\br{m_v^2-\pdiff{f\br{\avg{s},\avg{v}}}{\avg{v}}}^2}\avgabssq{\eta_x}}.
\end{equation}
We note that for positive autoregulation $\abs{J_{vv}}<\mu_v$ while for negative
autoregulation $\abs{J_{vv}}>\mu_v$. Thus the {\GNR } is larger for the positively autoregulated
than the three component cascade, especially for $\omega< J_{vv}$. For the negatively autoregulated
cascade the opposite holds.\\
The constraint does not lead to a unique relation between autoregulated and non-autoregulated
cascade. An alternative choice would be a simple three component cascade for which the
degradation rate $\mu_v$ is equivalent to the ''effective'' degradation rate in the autoregulated
cascade. Thus
$m_v=J^{ar}_{vv}$. The production of $x$ is then
\begin{equation}
\overbrace{\beta\frac{k_v}{\mu_v}}^{\text{autoregulated}}
=\overbrace{k_x\frac{k_v}{m_v}}^{\text{three component cascade}}
\end{equation}
Equalizing this leads to
\begin{eqnarray}
\beta\frac{k_v}{\mu_v}&=&k_x\frac{k_v}{m_v}\\
\to  k_x&=&\beta\frac{m_v}{\mu_v},
\end{eqnarray}
which leads to
\begin{equation}
\frac{g^2}{N}=\frac{\br{\beta\frac{\mu^{casc}_v}{\mu_v}k_v}^2}{\br{\beta\frac{m_v
} { \mu_v}}^2\avgabssq{\eta_v}
+\br{\omega^2+\br{m_v}^2}\avgabssq{\eta_x}}=\frac{\br{\beta
k_v}^2}{\beta^2\avgabssq{\eta_v}+\br{\frac{\mu_v}{m_v}}^2\br{
\omega^2+\br{m_v}^2}\avgabssq{\eta_x}}
\end{equation}
for the {\GNR } of the three component simple cascade. Since for positive feedback
$m_v<\mu_v$, the {\GNR} of the positively autoregulated cascade is larger than that of
the simple cascade, especially if $\omega\gg m_v$ or
$\omega^2\gg\beta^2\avgabssq{\eta_v}$.\\
If we allow for even more differences between the kinetic rates, but require equal
production, we obtain the following equations (we still
assume the signal to be identical in both cases)
\begin{equation}\label{eqn:si:constr2}
 \mu_v=C m_v \andd \beta = C k_x,
\end{equation}
where $C$ is an arbitrary constant.
We note that the mean level of $v$ differs between the autoregulated and the simple cascade
\begin{equation}\label{eqn:si:constr3}
\avg{v}^{ar}=\frac{1}{C}\avg{v}^{\rm simple}
\end{equation}
As a result we derive for the gain-to-noise ratio for the regulated cascade (using
Eqns.~\ref{eqn:si:gnr_ar_vs_simple},~\ref{eqn:si:constr2}~and~\ref{eqn:si:constr3})
\begin{eqnarray}
\frac{g^2}{N}&=&\frac{\br{Ck_xk_v}^2}{\br{Ck_x}^2\avgabssq{\eta_v}
+\br{\omega^2+\br{C\mu_v-\pdiff{f\br{\avg{s},\avg{v}}}{\avg{v}}}^2}\avgabssq{
\eta_x}} \nonumber \\
&=& \frac{\br{k_xk_v}^2}{\br{k_x}^2\avgabssq{\eta_v}
+\br{\frac{\omega^2}{C^2}+\br{\mu_v-\frac{1}{C}\pdiff{f\br{\avg{s},\avg{v}}}{
\avg{v}}}^2}\avgabssq{\eta_x}}
\end{eqnarray}
For small $\omega$ the conclusions on positive and negative feedback are
still valid, but for $\omega\to\infty$ the ratio of the {\GNR } for positive
feedback and a three component cascade is a function of $C$. Similar arguments can be
made about comparing negative and positive feedback for $\omega\to\infty$, where again the ratio of
the gain-to-noise ratio's depends on $C$.
\section{Feedback}
\subsection{Feedback from $x$ to $v$}
An elementary system with feedback from $x$ to $v$ is
\begin{subequations}
\begin{equation}
  \diff{v}{t}=f\br{x}s - \mu_v v+\eta_v\br{t}
\end{equation}
\begin{equation}
  \diff{x}{t}=\beta v -\mu_x x+\eta_x\br{t},
\end{equation}
\end{subequations}
where
\begin{equation} \label{eqn:si:kappa}
f\br{x}s=\frac{\nu \kappa^n s}{K^n+x^n}
\begin{cases}
  \kappa=K, & \text{negative feedback} \\
  \kappa=x, & \text{positive feedback}
\end{cases}
\end{equation}
For the gain, noise and $P_{XX}\br{\omega}$ we obtain
\begin{subequations}
\begin{equation}
g^2\br{\omega}=\frac{\br{J_{vs}\beta}^2}{\br{\omega^2+\mu_v^2}\br{
\omega^2+\mu_x^2}+J_{vx}\beta\sqbr{J_{vx}\beta+2\br{\omega^2-\mu_x\mu_v}}}
\end{equation}
\begin{equation}
N\br{\omega}=
\frac{\beta^2\avgabssq{\eta_v}+\br{\omega^2+\mu_v^2}\avgabssq{\eta_x}}{\br{
\omega^2+\mu_v^2}\br{\omega^2+\mu_x^2}+J_{vx}\beta\sqbr{J_{vx}\beta+2\br{
\omega^2-\mu_v\mu_x}}}
\end{equation}
\begin{equation}
P_{XX}\br{\omega}=\frac{\br{J_{vs}\beta}^2P_{SS}\br{\omega}+\beta^2\avgabssq{
\eta_v }
+\br{\omega^2+\mu_v^2}\avgabssq{\eta_x}}{\br{\omega^2+\mu_v^2}\br{\omega^2+\mu_x
^2}+J_{vx}\beta\sqbr{J_{vx}\beta+2\br{\omega^2-\mu_v\mu_x}}},
\end{equation}
\end{subequations}
where
\begin{equation}
J_{vx}=\pdiff{\ds \avg{f\br{x}s}}{\avg{x}}=\pdiff{\ds \frac{\nu \avg{\kappa}^n
\avg{s}}{K^n+\avg{x}^n}}{\ds \avg{x}}
\end{equation}
We note that the {\GNR} is independent of $J_{vx}$. The peak in $P_{XX}$ only exist if
$J_{vx}<0$, since
\begin{multline}
P_{XX}=\frac{\br{J_{vs}\beta}^2P_{SS}}{\br{\omega^2+\mu_v^2}\br{\omega^2+\mu_x
^2}+J_{vx}\beta\sqbr{J_{vx}\beta+2\br{\omega^2-\mu_v\mu_x}}} + \\
 +\frac{\beta^2\avgabssq{\eta_v }}{\br{\omega^2+\mu_v^2}\br{\omega^2+\mu_x
^2}+J_{vx}\beta\sqbr{J_{vx}\beta+2\br{\omega^2-\mu_v\mu_x}}} + \\
 +\frac{\mu_v^2\avgabssq{\eta_x}}{\br{\omega^2+\mu_v^2}\br{
\omega^2+\mu_x^2}+J_{vx}\beta\sqbr{J_{vx}\beta+2\br{\omega^2-\mu_v\mu_x}}} +\\
 +\frac{\omega^2\avgabssq{\eta_x}}{\br{\omega^2+\mu_v^2}\br{
\omega^2+\mu_x^2}+J_{vx}\beta\sqbr{J_{vx}\beta+2\br{\omega^2-\mu_v\mu_x}}}
\end{multline}
which are four monotonic decreasing functions of $\omega$ for $J_{vx}>0$.
So only for negative feedback a peak can exist in the power spectrum, gain and
noise (since the same argument applies to gain and noise).\\
The frequency of the maximum of the gain can easily be obtained, since it coincides with the minimum
of the denominator $D$
\begin{equation}
D=\br{\omega^2+\mu_v^2}\br{
\omega^2+\mu_x^2}+J_{vx}\beta\sqbr{J_{vx}\beta+2\br{\omega^2-\mu_x\mu_v}}.
\end{equation}
This frequency, where the gain has a maximum,  is
\begin{equation}
\omega^2_{\rm res}=-\frac{1}{2}\sqbr{\mu_v^2+\mu_x^2+2J_{xv}\beta},
\end{equation}
such that we require $\mu_v^2+\mu_x^2+2J_{xv}\beta<0$. As a check we note that $D>0$ for
$\omega_{\rm res}$ so divergence is not possible.
The maximum frequency for the noise is not the minimum of $D$, due to the $\omega$-dependence in the
numerator. If $\beta^2\avgabssq{\eta_v}\gg\avgabssq{\eta_x}$, the $\omega$-dependence in the
noise is less strong, and the frequency of the peak of the noise shifts to the frequency of the peak
in the gain.
Although a peak in $P_{XX}$ can be derived analytically ($\diff{P_{XX}}{\omega}$ is 4th order in
$\omega^2$), it is not insightful. We note that $P_{XX}$ is the sum of the noise ($N$) and the
signal ($\Sigma$), such that if one of these two dominates in $P_{XX}$ the peak is likely to
coalesce with the peak of the dominating term.
We also note that the signal $\Sigma$ depends on $\mu_s$, so the peak in $P_{XX}$ is not likely to
coincide exactly with the peak in the gain, since the gain is independent of $\mu_s$.\\
Compared with a three component cascade (rates $k_v, k_x$) , requiring equal production, we
note that
\begin{equation}
k_v=\frac{\nu \kappa^n}{K^n+\avg{x}^n}
\end{equation}
and the three component cascade has an identical {\GNR } as the cascade with
regulation.\\
\subsection{Linear Stability Analysis and Control theory}
We now shift gears and use some methods from linear stability analysis to study the
biochemical network from a slightly different perspective. After linearizing, the solution to the
linear differential equations for the perturbations is (ignoring the added noise)
\begin{subequations}
\begin{equation}
\diff{\vect{\tilde{y}}\br{t}}{t}=\mat{J}\vect{\tilde{y}}
\end{equation}
\begin{equation}
\vect{\tilde{y}}\br{t}=e^{\mat{J}t}\vect{\tilde{y}}\br{0}=
\mat{R}e^{\mat{\lambda}t}\mat{L}\vect{\tilde{y}}\br{0}
\end{equation}
\end{subequations}
Where $\mat{J}$ is the Jacobian, with eigenvalues $\lambda_i$ and right eigenvectors $r_i$.
The exponential matrix ($e^{\mat{J}t}$) describes the time dependency, and can decomposed in a
matrix
with diagonal entries $e^{\lambda_i}$, $\mat{R}$ with the right eigenvectors (as columns) and
$\mat{L}$ with the left eigenvectors (as rows). Alternatively, we could write down the solution in
terms of the right eigenvectors
\begin{equation}
\vect{\tilde{y}}\br{t}=c_1e^{\lambda_1t}\vect{r_1}+\ldots+c_ne^{\lambda_nt}\vect{r_n}
\end{equation}
where $c_1\ldots c_n$ are weighing coefficients which are obtained by solving
for the initial condition. In both expressions we note that the exponential exponent
involves $\lambda$. If $\lambda$ is complex, we can rewrite the exponent as
\begin{equation}
e^{\lambda t}=e^{\br{\Re_{\lambda}+i \Im_{\lambda}}
t}=e^{\Re_{\lambda}t}\br{\cos \br{\Im_{\lambda}t} + i
\sin\br{\Im_{\lambda}t} }
\end{equation}
and the fluctutations decay (if $\Re_{\lambda}<0$) with characteristic frequency
$\Im_{\lambda}$. For stability we require that $\mu_v\mu_x-J_{xv}J_{vx}>0$.\\
Yet another different method is control theory, which we can use to describe our system. In control
theory we describe a linear system using the convolution of a response function with the input to
determine the output.  In the fourier space this becomes multiplication, such that we have
(again ignoring noise)
\begin{subequations}
\begin{equation}
V\br{\omega}=H_1\br{\omega} S\br{\omega} + X_{fb}\br{\omega}
\end{equation}
\begin{equation}
X_{fb}\br{\omega} = G_1\br{\omega} X\br{\omega}
\end{equation}
\begin{equation}
X\br{\omega} = H_2\br{\omega} V\br{\omega}
\end{equation}
\end{subequations}
so that the total response function between input and output is
\begin{equation}
X\br{\omega}=\frac{H_1\br{\omega}H_2\br{\omega}}{1-H_2\br{\omega}G_1\br{\omega}}S\br{\omega}
\end{equation}
which is, if we take as transfer functions
\begin{equation}
H_1\br{\omega}=\frac{J_{vs}}{i\omega+\mu_v}\quad\quad H_2\br{\omega}=\frac{J_{xv}}{i\omega+\mu_x}
\quad\quad G_1\br{\omega}=\frac{\gamma}{i\omega+\mu_v}
\end{equation}
equal to $g\br{\omega}$.\\
The phase of the gain, which identifies the phase shift between $s$ and $x$ is
\begin{equation}
\Delta\phi=\arctan\br{\frac{-\omega\br{\mu_v+\mu_x}}{-\omega^2-J_{xv}\gamma+\mu_v\mu_x}}
\end{equation}
We can now define $\omega_\phi$
\begin{equation}
\omega_\phi=\sqrt{\mu_v\mu_x-J_{xv}\gamma}
\end{equation}
which defines the frequency for which the phase difference between $x$ and $s$ shifts by a factor
$\pi$. Since $x$ is also the fedback signal, this is the phase difference between the signals
in the feedback loop. For  negative feedback ($\gamma<0$) $\Delta\phi$ moves from $0$ to $\pi$
for $\omega$ changing from $0$ to $\infty$.

\subsection{Feedback from $w$ to $v$}
For the regulated four component cascade, the network is
\begin{subequations}
\begin{equation}
  \diff{v}{t}=\frac{\nu \kappa^n s}{K^n+w^n}- \mu_v  v+\eta_v\br{t}
\end{equation}
\begin{equation}
  \diff{w}{t}=\beta v -\mu_w  w+\eta_w\br{t}
\end{equation}
\begin{equation}
  \diff{x}{t}=\gamma w -\mu_x x+\eta_x\br{t},
\end{equation}
\end{subequations}
with $\kappa$ as before (e.g. Eqn.~\ref{eqn:si:kappa}). We linearize and obtain
\begin{subequations}
\begin{equation}
J_{vs}=\frac{\nu \kappa^n}{K^n+\avg{w}^n}
\end{equation}
\begin{equation} \label{eqn:si:jvw-3step}
J_{vw}=\pdiff{f\br{\avg{w}}}{\avg{w}}\avg{s}=-\frac{\nu
n\avg{w}^nK^n\avg{s}}{\avg{w}\br{K^n+\avg{w}^n}^2},
\end{equation}
\end{subequations}
where the equation \ref{eqn:si:jvw-3step} is for negative feedback (for positive
feedback the sign would be positive).
The gain and noise are
\begin{subequations}
\begin{equation}
g^2\br{\omega}=\frac{J_{vs}^2\beta^2\gamma^2}{\br{\omega^2+\mu_x^2}F\br{
\omega}}
\end{equation}
\begin{equation}
N\br{\omega}=\frac{\gamma^2\beta^2\avgabssq{\eta_v}+\gamma^2\br{
\omega^2+\mu_v^2}\avgabssq{\eta_w}+F\br{\omega}\avgabssq{\eta_x}}{\br{
\omega^2+\mu_x^2 }F\br{\omega}}
\end{equation}
\end{subequations}
and
\begin{equation}
F\br{\omega}=\omega^4+\br{\mu_{v}^2+\mu_{w}^2+2J_{vw}\beta}\omega^2+
\br{J_{vw}\beta-\mu_{w}\mu_{v}}^2,
\end{equation}
where $F\br{\omega}$ is a function of the parameters in the feedback loop only.
The {\GNR } is described by $C/a\br{\omega}$, i.e. a constant divided by a function of $\omega$.
For this to have an extremum, the denominator should have an extremum. We differentiate and obtain
\begin{equation} \label{eqn:si:wpeak}
\omega^2_{\rm peak}=-\frac{1}{2}\br{\gamma^2\frac{\avgabssq{\eta_w}}{\avgabssq{\eta_x}}
+\mu_v^2+\mu_w^2+2J_{vw}\beta}.
\end{equation}
Since this expression is negative, to have $\omega^2>0$ we require negative
feedback.
Explicitly writing $J_{wv}J_{vw}$, we have for the requirement that a peak exists
\begin{eqnarray}\label{si:eqn:nw}
&&2J_{wv}J_{vw}=2\frac{\nu
n\avg{w}^nK^n\avg{s}}{\avg{w}\br{K^n+\avg{w}^n}^2}\beta
>\br{\gamma^2\frac{\avgabssq{\eta_w}}{\avgabssq{\eta_x}}+\mu_v^2+\mu_w^2}
\\\avg{w}
&=&\frac{\beta\avg{v}} {\mu_w}=\frac{\beta}{\mu_v\mu_w}\frac{\nu
K^n\avg{s}} {K^n+\avg{w}^n}
\end{eqnarray}
which gives $n$ solutions for $\avg{w}$ (of which only one is real and positive). If we
constrain the production rate of $v$ and $w$  to be constant - and we assume
$\avg{v}=\frac{k_v\avg{s}}{m_v}$ - then we obtain
\begin{equation*}
\frac{\nu K^n}{K^n + \avg{w}^n}=k_v
\end{equation*}
and the following expression for Eqn.~\ref{si:eqn:nw}
\begin{equation}
\avg{w}=\frac{\beta\avg{v}}{\mu_w}=\frac{k_v\beta\avg{s}}{\mu_w\mu_v}.
\end{equation}
we rewrite the coupling strength $J_{vw}$
\begin{subequations} \label{eqn:SI:eqn_Jvw}
\begin{equation}
J^{\rm pos}_{vw}=\frac{\nu n
\avg{w}^{n-1}K^n\avg{s}}
{\br{K^n+\avg{w}^n}^2}=\frac{n\avg{s}k_v}{\avg{w}}\frac{K^n}{K^n+\avg
{w}^n}=\frac{n\avg{s}k_v}{\avg{w}}\frac{\br{\mu_w\mu_vK}^n}{\br{\mu_w\mu_vK}
^n+\br{k_vk_w\avg{s}}^n}
\end{equation}
\begin{equation}
J^{\rm neg}_{vw}=-\frac{\nu n \avg{w}^{n-1}
K^n\avg{s}} {\br{K^n+\avg{w}^n}^2}=-\frac{n\avg{s}k_v}{\avg{w}}\frac{\avg{w}^n}{
K^n+\avg{w}^n}
=-\frac{n\avg{s}k_v}{\avg{w}}\frac{\br{k_vk_w\avg{s}}^n}{\br{
\mu_w\mu_vK}^n+\br{k_vk_w\avg{s}}^n}
\end{equation}
\end{subequations}
For $K\ll\avg{w}$ positive regulation is maximized and $J^{\rm  pos}_{vw}$ is maximal, while
negative regulation is greatly suppressed and $\abs{J_{vw}^{\rm neg}}$ is minimal. The limit
$n\to\infty$ is more complicated. If $K < \avg{w}$, $J^{\rm neg}_{vw}\to -\infty$, while $J^{\rm
pos}_{vw}\to\infty$ for $\avg{w} < K$. In the opposite scenario's the limits tend to zero. This is
only valid if while changing $n$, $\avg{w}$ remains constant, which is true due the constraint.\\
With Eqn.~\ref{eqn:SI:eqn_Jvw} we can study $\omega_{\rm peak}$ in more detail and we obtain
\begin{eqnarray}
2\frac{\nu n\avg{w}^nK^n\avg{s}}{\avg{w}\br{K^n+\avg{w}^n}^2}\beta
&>&\br{\gamma\mu_w+\mu_v^2+\mu_w^2}\nonumber\\
\frac{2k_v\beta
n\avg{s}}{\avg{w}}\frac{\avg{w}^n}{K^n+\avg{w}^n}&>&\br{
\gamma\mu_w+\mu_v^2+\mu_w^2}\nonumber\\
2\frac{\avg{w}^n}{K^n+\avg{w}^n}\mu_v\mu_w n&>&\br{\gamma\mu_w+\mu_v^2+\mu_w^2}
\end{eqnarray}
which, interestingly, only has a solution for $n>1$.\\
The power spectrum of $x$ is
\begin{equation}
P_{XX}\br{\omega}=\frac{J_{vs}^2\beta^2\gamma^2P_{SS}+\gamma^2\beta^2\avgabssq{\eta_v}
+\gamma^2\br{\omega^2+\mu_v^2}\avgabssq{\eta_w}+F\br{\omega}\avgabssq{
\eta_x}}{\br{\omega^2+\mu_x^2}F\br{\omega}},
\end{equation}
which depends on $\mu_s$ through $P_{SS}$ and therefor will have a peak for a different $\omega$
than the \GNR.\\
The {\GNR } for the simple four component cascade is
\begin{equation}
\frac{g^2\br{\omega}}{N\br{\omega}}=\frac{\br{k_vk_wk_x}^2}{k_x^2\sqbr{k_w^2\avgabssq{
\eta_v}+\br{\omega^2+m_v^2}\avgabssq{\eta_w}}+\br{\omega^2+m_v^2}\br{\omega^2+m_w^2}
\avgabssq{\eta_x}}=\frac{\br{k_vk_wk_x}^2}{D}
\end{equation}
where we chose $\nu$ such that
\begin{equation}
k_v=\frac{\nu \kappa^n}{K^n+\avg{w}^n}
\end{equation}
to obtain equal production. We then  obtain for the ratio of the
{\GNR } of the feedback cascade and a simple cascade
\begin{equation}
G=\frac{GNR_{\rm fb}}{GNR_{\rm simple}}=\frac{D}{D+J_{vw}\beta\sqbr{J_{vw}\beta+2\br{
\omega^2-\mu_v\mu_w}}\avgabssq{\eta_x}}
\end{equation}
So that the feedback is larger if
$J_{vw}\beta\sqbr{J_{vw}\beta+2\br{\omega^2-\mu_v\mu_w}}\avgabssq{\eta_x}<0$.\\
The result of this inequality is
\begin{eqnarray}
	G_{\rm pos}(\omega) >1 &\text{ if }&
		\omega^2<\mu_v\mu_w\br{1-\frac{n}{2}\frac{K^n}{K^n+\avg{w}^n}}, \\
	G_{\rm neg}(\omega) >1 &\text{ if}&
		\omega^2>\mu_v\mu_w\br{1+\frac{n}{2}\frac{\avg{w}^n}{K^n+\avg{w}^n}}.
\end{eqnarray}
which are Eqns.~\ref{eqn:fb_omega_switch} from the article.
The peak for the negative feedback occurs at $\omega_{\rm peak}$ (Eqn.~\ref{eqn:si:wpeak}).
The negative feedback cascade is larger than the four component simple cascade if
$\omega>\omega_{\rm
switch}$ (Eqns.~\ref{eqn:fb_omega_switch}). Thus if $\omega_{\rm peak}>\omega_{\rm switch}$ the
{\GNR } for the negative feedback at the peak is larger than the four component cascade
\begin{eqnarray}
\omega_{\rm peak}^2&>&\mu_v\mu_w\sqbr{1+\frac{n}{2}\frac{\br{k_v\beta\avg{s}}^n}{\br{
\mu_v\mu_wK }^n+\br{k_v\beta\avg{s}}^n}}\nonumber\\
\br{\gamma^2\frac{\avgabssq{\eta_w}}{\avgabssq{\eta_x}}
+\mu_v^2+\mu_w^2+2J_{vw}\beta}&<&-\mu_v\mu_w\sqbr{2+n\frac{\br{
k_v\beta\avg{s}}^n}{\br{\mu_v\mu_wK}^n+\br{k_v\beta\avg{s}}^n}}\nonumber\\
n\mu_v\mu_wM&>&\br{\mu_v+\mu_w}^2+2\gamma^2\frac{\avgabssq{\eta_w}}{\avgabssq{\eta_x}}
\end{eqnarray}
which is possible for large $n$ and large $M=\frac{\avg{w}^n}{\avg{w}^n+K^n}$, which indicates that
$K\ll\avg{w}$, in both cases representing a strong negative feedback.\\

\section{Comments on Fig.~\ref{fig:fb_int}E}
Here we list some additional explanation on Fig.~\ref{fig:fb_int}E. In this figure, we keep the
parameters $\mu_v$, $\mu_w$, $\nu$, $\beta\br{=J_{wv}}$ and $K$ constant, since they dictate
the feedback cycle (Eqn.~1 in Fig.~\ref{fig:fb_int}B). We vary $J_{xw}$ and $\mu_x$, so
that in this case, not the average production rate of $x$ is constrained, but the average copy
number $\avg{x}$.\\
To understand the dependence of the gain, noise and gain-to-noise ratio on $\gamma=J_{xw}$ and
$\mu_x$,we note that $g^2\sim\gamma^2g^2_{s\to w} /(\mu_x^2+\omega^2)$ and
$N\sim\gamma^2/(\mu_x^2+\omega^2)N_{v}(\omega)+\gamma^2/(\mu_x^2+\omega^2)N_{w}(\omega)+N_x(\omega)$
, where $N_{v}(\omega)$ and $N_{w}(\omega)$ are independent of $\gamma$ and $\mu_x$
and $N_{x}(\omega)=2\gamma\avg{w}/(\mu_x^2+\omega^2)$ (with $\avg{w}$ being independent of
$\gamma$ and $\mu_x$).\\

For $\omega \ll \mu_x$, the contributions of $v$ and $w$ to $N(\omega)$
are proportional to $\gamma^2/\mu_x^2$, while the contribution of $x$ is
given by $N_x(\omega) \propto \gamma / \mu_x^2$. Hence, for $\omega \ll
\mu_x$, the contributions of $v$ and $w$ to the noise are constant,
while the contribution of $x$ decreases with increasing $\gamma$ and
$\mu_x$, leading to a decrease of $N(\omega)$. Since the gain is
constant in this regime, the gain-to-noise ratio increases with
increasing $\gamma$ and $\mu$ for $\omega \ll \mu_x$. For $\omega \gg
\mu_x$, the gain, and the contributions of $v$ and $w$ to the noise
increase with $\gamma^2$ while the contribution of $x$ to the noise
increases with $\gamma$, meaning that also in this regime the
gain-to-noise ratio increases with $\gamma$ and $\mu_x$.

\end{document}